\documentclass[12pt,a4paper]{article}
\pdfoutput=1
\usepackage{jcappub}
\usepackage{graphicx}
\usepackage{amssymb}
\usepackage{amsmath}
\usepackage{bm}
\usepackage{hyperref}

\graphicspath{{figures/}}

\begin{document}
\title{The evidence of cosmic acceleration and observational constraints}

\author[a]{Yingjie Yang,}

\author[a,1]{Yungui Gong\note{Corresponding author.},}

\affiliation[a]{School of Physics, Huazhong University of Science and Technology,
Wuhan, Hubei 430074, China}

\emailAdd{yyj@hust.edu.cn}
\emailAdd{yggong@hust.edu.cn}

\abstract{
Directly comparing the 6 expansion rate measured by type Ia supernovae data and the
lower bound on the expansion rate set by the strong energy conditions
or the null hypothesis that there never exists cosmic acceleration,
we see $3\sigma$ direct evidence of cosmic acceleration and
the $Rh=ct$ model is strongly excluded by the type Ia supernovae data.
We also use Gaussian process method to reconstruct the expansion rate
and the deceleration parameter from the 31 cosmic chronometers data
and the 6 data points on the expansion rate measured from type Ia supernoave data,
the direct evidence of cosmic acceleration is more than $3\sigma$
and we find that the transition redshift $z_t=0.60_{-0.12}^{+0.21}$
at which the expansion of the Universe
underwent the transition from acceleration to deceleration.
The Hubble constant inferred from the cosmic chronometers data with
the Gaussian process method is $H_0=67.46\pm4.75$ Km/s/Mpc.
To understand the properties of cosmic acceleration and dark energy, we fit two different two-parameter models to the observational data,
and we find that the constraints on the model parameters from
either the full distance modulus data by the Pantheon compilation
or the compressed expansion rate data are very similar,
and the derived Hubble constants are consistent with the Planck 2018 result.
Our results confirm that the 6 compressed expansion rate data can replace
the full 1048 distance modulus data from the Pantheon compilation.
We derive the transition redshift $z_t=0.61_{-0.16}^{+0.24}$
by fitting a simple $q(z)$ model to the combination of cosmic chronometers data
and the Pantheon compilation, the result is consistent
with that obtained from the reconstruction with Gaussian process.
By fitting the observational data by the SSLCPL model which approximates the dynamics of general
thawing scalar fields over a large redshift range,
we obtain that $H_0=66.8\pm 1.4$, $\Omega_{\phi 0}=0.69\pm 0.01$ and $w_0=-1.03\pm 0.07$. The result shows that $\Lambda$CDM model is consistent with the observational data.}

\arxivnumber{1912.07375}

\maketitle

\section{Introduction}

The observations of type Ia supernovae (SNe Ia) \cite{Riess:1998cb,Perlmutter:1998np} suggest that the Universe is currently undergoing accelerated expansion.
This raises a vital question about the mechanism of this accelerated expansion: what is the cause and nature of the accelerated expansion?
Many efforts have been made to understand this question and two approaches
were usually used. One approach is to modify general relativity at the
cosmological scale, such as the Dvali-Gabadadze-Porrati model \cite{Dvali:2000hr},
f(R) gravity \cite{Carroll:2003wy,Nojiri:2003ft,Starobinsky:2007hu,Hu:2007nk}, and
dRGT ghost-free massive gravity \cite{deRham:2010kj,Gong:2012yv}.
The other approach is to introduce an exotic matter component dubbed as dark energy
which has negative pressure and contributes about 70\% of
the matter content of the Universe to drive the cosmic acceleration.
Although the cosmological constant named as $\Lambda$CDM model is the
simplest candidate for dark energy and is consistent with current
observations, it also faces problems such as fine tuning and coincidence problems.
Furthermore, there exist many orders of magnitude discrepancy
between the theoretical estimation
and astronomical observations for the cosmological constant \cite{Weinberg:1988cp}.
Therefore, dynamical dark energy models such as the quintessence model
\cite{Ratra:1987rm,Wetterich:1987fm,Caldwell:1997ii,Zlatev:1998tr,Steinhardt:1999nw} are usually considered.
For a recent review of dark energy, please see refs. \cite{Sahni:1999gb,Copeland:2006wr,Padmanabhan:2007xy,Li:2011sd,Benetti:2019gmo}.

The Hubble constant $H_0=67.27\pm 0.60$ km/s/Mpc inferred from Planck 2018
measurement on the cosmic microwave background anisotropy (CMBR) with
the assumption of $\Lambda$CDM model \cite{Aghanim:2018eyx}
is in $4.4 \sigma$ tension with the local measurement $H_0=74.03\pm 1.42$ km/s/Mpc
by the Hubble Space Telescope (HST) observations of 70 long-period Cepheids in the large Magellanic Cloud \cite{Riess:2019cxk}. Combining the distance measurement
from gravitational wave and the identification of local host galaxy
from the electromagnetic counterpart, gravitational wave becomes a standard
siren and can be used to measure the Hubble constant \cite{Schutz:1986gp}.
The detection of the first gravitational wave event GW170817 and its
electromagnetic counterpart GRB170817A from a binary
neutron star merger measures $H_0$ as $H_0=70.0^{+12.0}_{-8.0}$ km/s/Mpc \cite{Abbott:2017xzu}. This value is consistent with both local
and high redshift measurements due to the large error bar.
By reconstructing the observational data of the Hubble parameter $H(z)$
from cosmic chronometers (CCH) and baryon acoustic oscillation (BAO)
with Gaussian process (GP) method, it was found that $H_0\sim 67\pm 4$
km/s/Mpc \cite{Yu:2017iju}. Using the Gaussian kernel in the GP method,
the reconstruction of CCH data and SNe Ia data from the Pantheon
compilation \cite{Scolnic:2017caz} and the HST CANDELS and
CLASH Multi-Cycle Treasury (MCT) programs
\cite{Riess:2017lxs} (Patheon+MCT) gives $H_0=67.06\pm 1.68$ km/s/Mpc \cite{Gomez-Valent:2018hwc}.
Applying the GP method with the Mat\'{e}rn kernel
to the combination of CCH, BAO and SNe Ia data,
it was found that $H_0=68.52^{+0.94+2.51(\text{sys})}_{-0.94}$ km/s/Mpc \cite{Haridasu:2018gqm}.
The results with different kernels are consistent with each other.
These values prefer the lower value determined from Planck 2018 data
and is in tension with the local measurement from distance ladder.

The tension on the Hubble constant may not caused by the fitting model \cite{Gao:2013pfa}. To check the tensions in data,
null tests with the reconstruction of some smooth functions
from observational data  may be used \cite{Clarkson:2007pz,Sahni:2008xx,Zunckel:2008ti,Nesseris:2010ep,Shafieloo:2012rs,Yahya:2013xma,Nesseris:2014mfa,Marra:2017pst,Franco:2019wbj,Sahni:2014ooa}.
The GP method is one of the most widely used model independent method to reconstruct a function and its derivatives from discrete data points without invoking any specific model. This method has been used in cosmology to reconstruct cosmological
parameters and probe the property of cosmic acceleration \cite{Clarkson:2010bm,Shafieloo:2012ht,Holsclaw:2010nb,Holsclaw:2010sk,Holsclaw:2011wi,Bilicki:2012ub,Seikel:2012uu,Seikel:2012cs,Seikel:2013fda,Nair:2013sna,Busti:2014dua,Sahni:2014ooa,Verde:2014qea,Li:2015nta,Vitenti:2015aaa,Wang:2016iij,Zhang:2016tto,Wei:2016xti,Yu:2017iju,Yennapureddy:2017vvb,Melia:2018tzi,Gomez-Valent:2018hwc,Pinho:2018unz,Haridasu:2018gqm,Jesus:2019nnk,Bengaly:2019ibu}.

The evidence for cosmic acceleration and the measurement on the Hubble constant
from Planck 2018 data were obtained by fitting the observational data,
so they depend on the models used in the fitting. The zero acceleration
model (eternal coasting \cite{John:1999gm} or $Rh=ct$ model \cite{Melia:2007sd}) is also consistent with some observational data \cite{Melia:2011fj,Lopez-Corredoira:2016pwg,Melia:2016djn,Melia:2018nfw,John:2019nlw,Yennapureddy:2019omi}.
To be model independent, many parametric and none-parametric model independent
(in the sense that it does not use a particular cosmological model) methods \cite{Capozziello:2018jya,Arjona:2019fwb} were proposed to study the evolution of the deceleration parameter $q(z)$, the geometry of the Universe and the property of dark energy.
In particular,
by comparing the bound set by the null hypothesis that the Universe never
experiences an accelerated expansion with the observational data,
the energy conditions may be used to provide direct and model independent
evidence of cosmic acceleration \cite{Visser:1997qk,Visser:1997tq,Santos:2006ja,Santos:2007pp,Gong:2007fm,Gong:2007zf,Seikel:2007pk,Velten:2017ire}.
As emphasized in \cite{Gong:2007fm,Gong:2007zf},
great caution is needed to correctly interpret the result from
falsifying the null hypothesis. The violation of the bound set by the null hypothesis
provides direct evidence that cosmic acceleration once occurred,
and the fulfillment of the bound doesn't mean no cosmic acceleration at all,
which is the reason why no evidence of acceleration was found in accelerating cosmologies in ref. \cite{Velten:2017ire}.
Therefore, the bound set by the null hypothesis or energy conditions
provides the direct model-independent evidence of cosmic acceleration if we interpret the result correctly.
In this paper, we
analyze the direct evidence of cosmic acceleration using the CCH data
and the expansion rate data $E(z)$ from the
Pantheon+MCT SNe Ia compilation \cite{Riess:2017lxs}.
The evidence with the CCH data depends on the Hubble constant $H_0$ which leads to model dependence, so the expansion rate data $E(z)$ provides stronger evidence in a model independent way.
However, there are only six $E(z)$ data points, so we use the GP method to reconstruct the expansion rate $E(z)$
and the deceleration parameter $q(z)$ from the CCH and Pantheon+MCT data.

Although the kinematic method does not assume any
gravitational theory and matter content, it only addresses
the question whether the Universe once experienced accelerated expansion and it cannot provide us any detailed information about
the cosmic acceleration, like the transition redshift at which
the expansion of the Universe underwent the transition
from accelerated expansion to decelerated expansion.
To probe the properties of cosmic acceleration and dark energy with the combination of
different observational data, we parameterize the deceleration parameter $q(z)$ with a simple two-parameter model $q(z)=1/2+(q_1z+q_2)/(1+z)^2$ \cite{gong:2006tx,gong:2006gs} and the equation of state parameter $w(z)$ by the
SSLCPL model \cite{Gao:2012ef,Gong:2013bn} which approximates the dynamics of general
thawing scalar fields over a large redshift range.

The paper is organized as follows. In section \ref{sec2},
we discuss the null hypothesis method and the
direct evidences for cosmic acceleration from Pantheon+MCT data and CCH data. In section \ref{sec3}, we use the GP method to reconstruct the expansion rate and deceleration parameter
by combining the CCH and Pantheon+MCT SNe Ia data.
To understand the properties of cosmic acceleration and dark energy better, two particular parameterizations are used to fit the observational
data in section \ref{sec4}.
We conclude the paper with some discussions in section \ref{sec5}.
The observational data and the GP method are presented in appendices \ref{appdxa} and \ref{appdxb}.

\section{Direct evidence for cosmic acceleration}
\label{sec2}

In this section, we start from the conditions
\begin{gather}
\label{secqt}
  q(t)\equiv -\frac{\ddot{a}}{aH^2}\geq 0, \\
\label{secht}
 \dot{H}-\frac{k}{a^2}\leq 0.
\end{gather}
The condition \eqref{secqt} means no acceleration
and the condition \eqref{secht} means no super-acceleration,
these conditions are also called null hypothesis.
The model with $q=0$
is also called eternal coasting \cite{John:1999gm} or $Rh=ct$ model \cite{Melia:2007sd}. Integrating equation \eqref{secht} yields
\begin{equation}
\label{sechz1}
  H(z)\geq H_0\sqrt{1-\Omega_k+\Omega_k(1+z)^2},
\end{equation}
where the Hubble constant $H_0$ denotes the current value of the Hubble parameter $H(z)$ and $\Omega_k=-k/(a_0 H_0^2)$. For a spatially flat universe, $\Omega_k=0$,
the above condition becomes $H(z)\ge H_0$.

By using the redshift $z$, the deceleration parameter $q(t)$ is related
with the Hubble parameter $H(t)$ as
\begin{equation}
\label{hzqzeq}
  \ln\frac{H(z)}{H_0}=\int_{0}^{z}\frac{1+q(z')}{1+z'}dz'.
\end{equation}
Substituting equation \eqref{secqt} into equation \eqref{hzqzeq}, we get
\begin{equation}
\label{sechzeq2}
  H(z)\geq H_0(1+z).
\end{equation}
If the universe has never experienced an accelerated
expansion or the expansion is always decelerating,
then equation \eqref{sechzeq2} is always satisfied.
Therefore, this simple argument can be used to obtain direct model-independent
evidence for cosmic acceleration.
However, we must be cautious to interpret the result correctly.
Because of the integration effect, even if the condition \eqref{sechzeq2} is satisfied
at some redshifts,
it does not mean that the universe has never experienced an accelerating
expansion \cite{Gong:2007fm,Gong:2007zf}.
If the condition \eqref{sechzeq2}
is violated at some redshifts, we are sure that the universe once experienced accelerating expansion.

For $z\ge 0$, we have
\begin{equation}
\label{sechzeq3}
  H_0(1+z)\geq H_0\sqrt{1-\Omega_k+\Omega_k(1+z)^2}.
\end{equation}
Therefore, once the universe experiences super-accelerated expansion, it must also experience accelerating expansion.
If we assume Friedmann-Robertson-Walker (FRW) metric and Einstein's general relativity, then the conditions \eqref{secqt} and \eqref{secht} can be derived from the strong energy conditions $\rho+3p\ge 0$ and $\rho+p\ge 0$ by using the Friedmann equation,
and Eq. \eqref{sechzeq3} tells us
that once the energy condition $\rho+3p\ge 0$ is satisfied,
the condition $\rho+p\ge 0$ is also satisfied.
Although the conditions \eqref{sechz1}
and \eqref{sechzeq2} can be derived from the strong energy conditions in the standard cosmological framework, these bounds actually are independent of the strong energy conditions and can be applied to more general cases because they just depend
on the conditions \eqref{secqt} and \eqref{secht}
and the FRW metric for the physical interpretation
of the scale factor $a(t)$. In other words,
the lower bound \eqref{sechzeq2} for decelerated expansion is
independent of a particular theory of gravity such as general relativity and it just assumes the FRW metric.

We can compare the conditions \eqref{sechz1}
and \eqref{sechzeq2} with observational data to show direct evidence
of both accelerated and super-accelerated expansion,
so the CCH data on $H(z)$ and the SNe Ia data on $E(z)$ can be used to see whether
the Universe ever experienced accelerated expansion or not, i.e.,
we can compare equations \eqref{sechz1} and \eqref{sechzeq2} with observational data $H(z)$ or $E(z)$ to
show direct evidence of cosmic acceleration in a model independent way. For the comparison with $E(z)=1+z$, it is totally free of any cosmological parameter. For the comparison with $H(z)=H_0(1+z)$,
it depends on the value of the Hubble constant, so the evidence becomes weaker.

\begin{figure}[htb]
\centering
\includegraphics[width=0.6\linewidth]{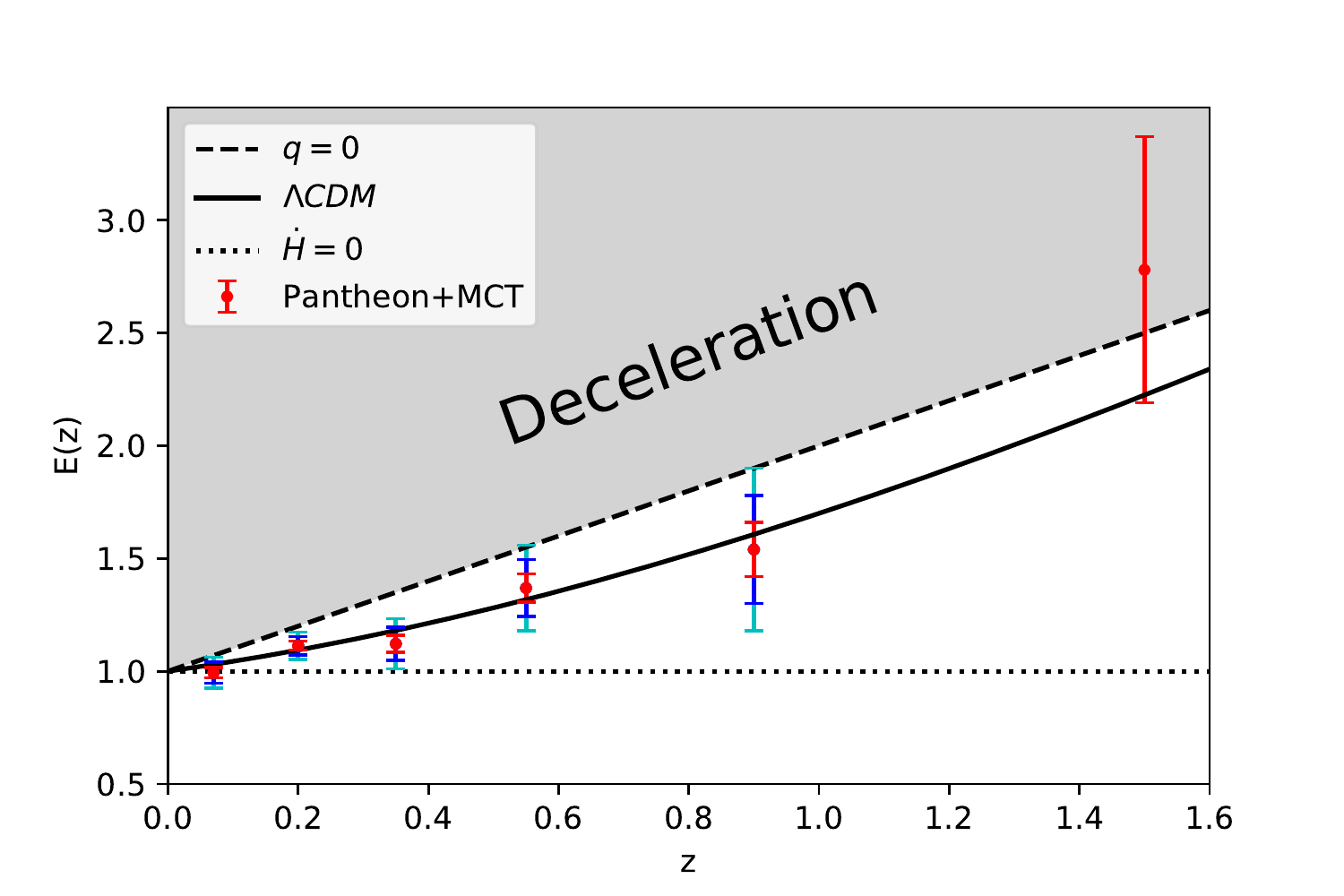}
\caption{The Pantheon+MCT SNe Ia measurements on $E(z)$ with $1\sigma$,
$2 \sigma$ and $3\sigma$ errors. The dashed line corresponds to the $Rh=ct$
model with $q(z)=0$, the dotted line denotes $E(z)=1$ which represents the
model with $\dot{H}=0$ in a spatially flat universe, and the solid line shows
the best fit $\Lambda$CDM model.}
\label{ezfig1}
\end{figure}

Now we use the Pantheon+MCT SNe Ia measurements on $E(z)$
to show the direct evidence for cosmic acceleration. The advantage of
the $E(z)$ data \cite{Riess:2017lxs} is that it is independent of the Hubble constant and the
drawback is that it assume $\Omega_k=0$,
so it is model dependent in this sense.
For the compressed SNe Ia data at six
redshifts, we compare $E(z)$ data with $1+z$ and $1$ to show the evidence of
accelerated expansion or super-accelerated expansion, respectively.
We plot the $E(z)$ data from table \ref{eztable} in appendix \ref{appdxa} along
with the null hypotheses \eqref{sechz1}
and \eqref{sechzeq2} in figure \ref{ezfig1}.
From figure \ref{ezfig1}, we see that all the low redshift points ($z<1$)
violate the lower bound \eqref{sechzeq2} even at $3\sigma$ level,
so we have $3\sigma$ evidence for cosmic acceleration, but we don't see
strong evidence for decelerated expansion due to
the lack and poor quality of data at high redshifts and the integration effect in the bound.
The first point also provides weak evidence of super-accelerated expansion.
Therefore, the Pantheon+MCT SNe Ia data is
strongly against the $Rh=ct$ model.
Note that this does not mean that the expansion
of the Universe is always accelerating up
to the redshft $z\sim 1$ or there is no decelerated expansion
at all as we explained above.
The $E(z)$ graph just provides us with the evidence
for cosmic acceleration and it does not give us any
information about the property of cosmic acceleration,
we will discuss the properties of cosmic acceleration and dark energy below. This evidence is independent of any gravitational theory
and it just assumes the flat FRW metric. Fitting $Rh=ct$ model to
Pantheon+MCT data, we get $\chi^2=85.29$. For $\Lambda$CDM model, the best fit
is $\Omega_{m0}=0.265\pm 0.029$ and $\chi^2=7.69$, so $Rh=ct$ model is strongly
disfavoured by the Pantheon+MCT data.

\begin{figure}[htbp]
\centering
\includegraphics[width=0.6\linewidth]{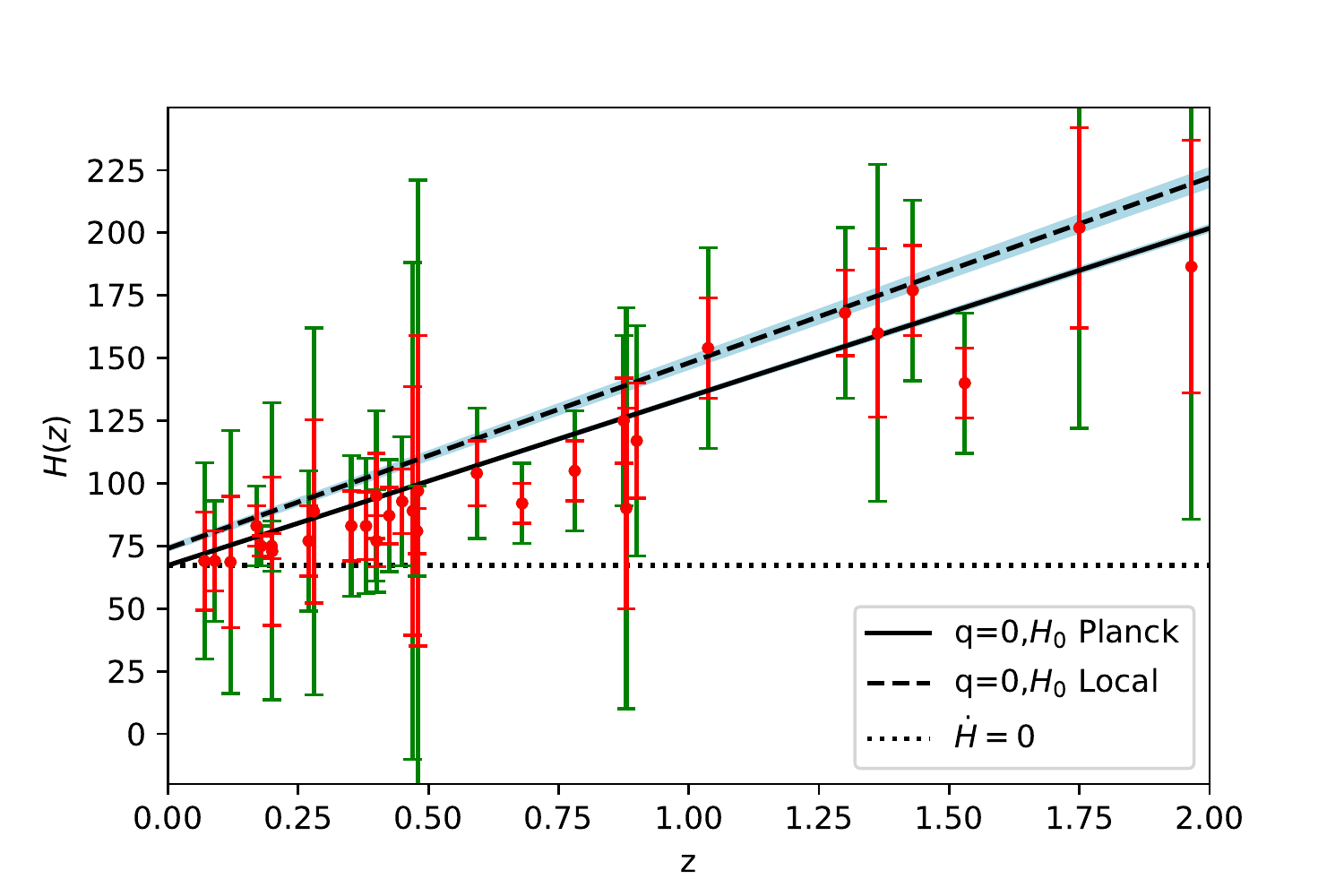}
\caption{The CCH data with $1\sigma$ and 2$\sigma$ uncertainties. The solid lines corresponds to $q(z)=0$ with $H_0=67.27$ km/s/Mpc and the blue shaded area corresponds to $H_0=67.27\pm 0.60$ km/s/Mpc. The dotted line denotes $H(z)=H_0=67.27$ km/s/Mpc.
The dashed line corresponds to $q(z)=0$ with $H_0=74.03$ km/s/Mpc and the blue shaded area corresponds to $H_0=74.03\pm1.42$ km/s/Mpc.}
\label{hzfig1}
\end{figure}

Then we compare the CCH measurements on the Hubble parameter $H(z)$
with the null hypothesis \eqref{sechz1}
and \eqref{sechzeq2} and the result is shown in figure \ref{hzfig1}.
Since the Hubble constant $H_0$ appears in equations \eqref{sechz1}
and \eqref{sechzeq2},
and the latest result $H_0=67.27\pm 0.60$ km/s/Mpc from Planck 2018 \cite{Aghanim:2018eyx}
is in tension with the local measurement $H_0=74.03\pm 1.42$ km/s/Mpc from HST \cite{Riess:2019cxk} at $4.4 \sigma$ level, so the direct evidence of cosmic acceleration from CCH data
is affected by the uncertainty in $H_0$ and this method
depends on the cosmological model as the way
the Hubble constant depends on. In figure \ref{hzfig1}, we take both the values
$H_0=67.27\pm 0.60$ km/s/Mpc and  $H_0=74.03\pm 1.42$ km/s/Mpc.
For $H_0=74.03$ km/s/Mpc, some of the low redshift $(z<1)$ CCH data violate the lower bound of the null hypothesis \eqref{sechzeq2}, and the three points at
$z=0.68$, $z=0.781$ and $z=1.53$ violate the lower bound even
at $2\sigma$ confidence level. So we expect that the Universe once experienced accelerated expansion, but this does not mean that the accelerated expansion happened
up to $z=1.53$. For $H_0=67.27$ km/s/Mpc, the data points that violate the
lower bound \eqref{sechzeq2} become less and only two points at $z=0.68$
and $z=1.53$ violate the lower bound at $2\sigma$ confidence level. Therefore,
the evidence strongly depends on the value of $H_0$. For both choices of $H_0$,
we see $2\sigma$ evidence of cosmic acceleration and no strong evidence of super-accelerated expansion,
note that this result applies to any value of $\Omega_k$.
Comparing with the evidence from $E(z)$ data, this evidence is much
weaker and it depends on the value of $H_0$ as discussed above.
On the other hand, we may argue that those points which violate the lower bound are outliers, so we fit
the $Rh=ct$ model to the CCH data and we get
the best fit $H_0=62.34\pm 1.43$ km/s/Mpc and $\chi^2=16.62$.
For the $\Lambda$CDM model, the best fit is
$\Omega_{m0}=0.32\pm0.06$, $H_0=68.11\pm 3.09$ km/s/Mpc
and $\chi^2=14.50$. In terms of $\chi^2$ statistics, it seems that both
$Rh=ct$ and $\Lambda$CDM model fit CCH data well.
To avoid the possible outlier problem,
in the next section we reconstruct the data using the GP method.

\section{Reconstruction of observational data with Gaussian process}
\label{sec3}

\begin{figure}[htbp]
\centering
\includegraphics[width=0.6\linewidth]{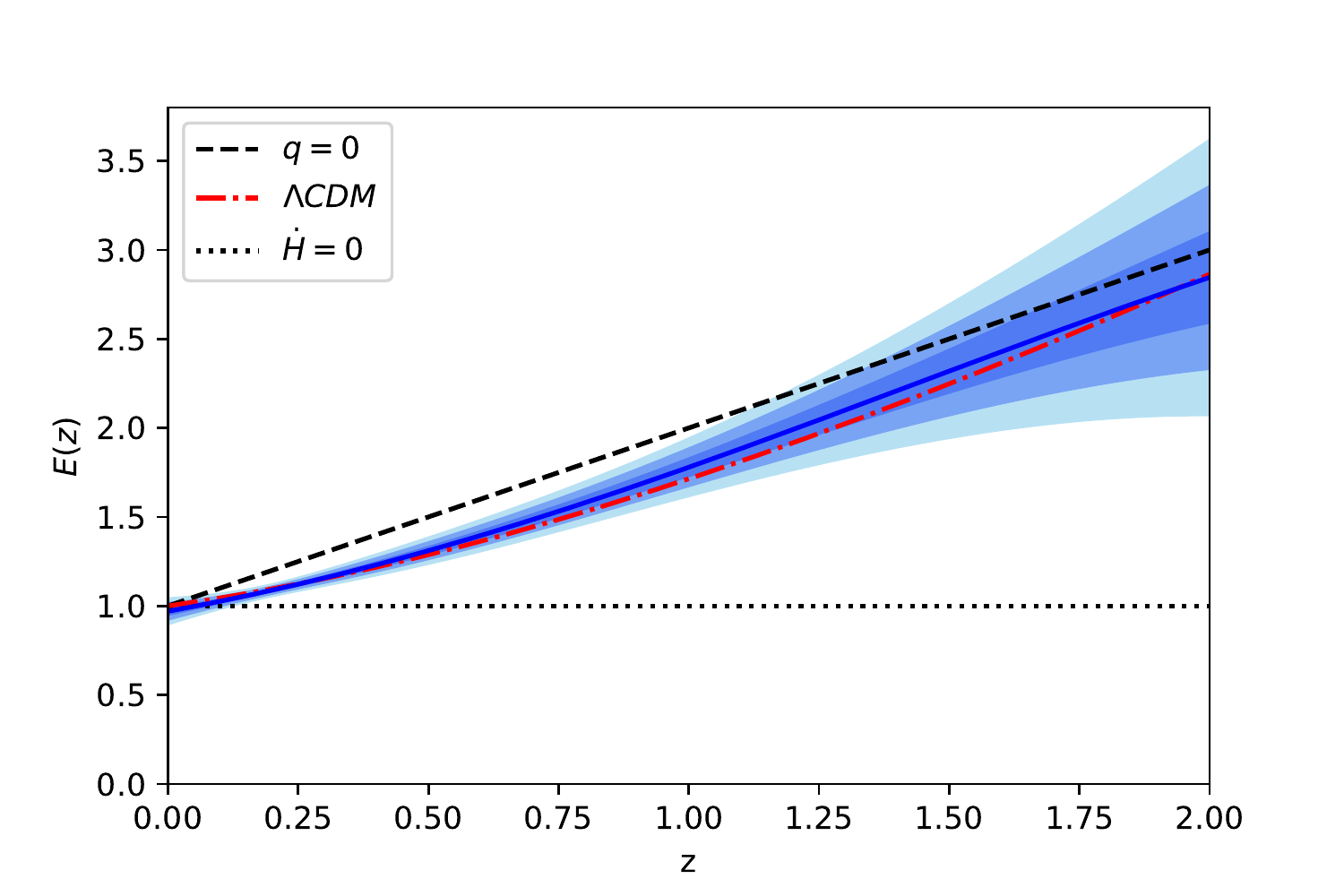}
\caption{GP reconstruction of $E(Z)$ from CCH+Patheon+MCT data.
The blue solid line is the mean of the reconstruction and the shaded areas are $1\sigma$, $2\sigma$ and $3\sigma$ errors. The dashed line corresponds
to $q(z)=0$ ($E(z)=1+z$) and the dotted line corresponds to $E(z)=1$.}
\label{gpezfig}
\end{figure}

Although we conclude that the Universe once experienced accelerated expansion just by comparing $E(z)$ data with the bound $1+z$,
we have no idea on the property of cosmic acceleration like when the cosmic acceleration began.
In this section, we use the GP method to reconstruct $E(z)$ and $q(z)$. The detailed discussion of GP method is presented in appendix \ref{appdxb}.
Since there are only six $E(z)$ data, so we reconstruct $E(z)$ by combining the CCH data on the Hubble parameter and the Pantheon+MCT data on $E(z)$ to show direct evidence of cosmic acceleration.
As seen from equation \eqref{sechzeq2}, to derive $E(z)=H(z)/H_0$ from $H(z)$,
we need to determine the Hubble constant $H_0$ from
the reconstructed $H(z)$ function at $z=0$ by using the
31 CCH data. We use the inferred value of $H_0=67.46\pm4.75$ km/s/Mpc from the GP reconstruction of the CCH data and divide the CCH data by this $H_0$ to obtain
$E(z)$ from the CCH data, then we add these $E(z)$ data to the Pantheon+MCT data
to reconstruct $E(z)$ function.
The result is shown in figure \ref{gpezfig}.
The reconstructed value $H_0=67.46\pm4.75$ km/s/Mpc is consistent
with the result $H_0\sim 67\pm 4$ km/s/Mpc obtained
from the 31 CCH data and 5 BAO data with the GP method in \cite{Yu:2017iju}. It is also consistent with reconstructed result
$H_0=67.06\pm 1.68$ km/s/Mpc obtained
from the combination of CCH and SNe Ia data in \cite{Gomez-Valent:2018hwc}.
Because the reconstruction starts from $z=0$ and $E(z=0)=1$ by definition, we expect the convergence at $E(z=0)=1$
due to the Hubble law $H(z)\approx H_0$ at low redshift which is independent of cosmological models to the lowest order.
However, the large uncertainties in $H(z)$
decrease the constraint ability of this method near $z=0$.
From figure \ref{gpezfig}, we see $3\sigma$ evidence of accelerated expansion in
the redshift ranges $0.1\lesssim z\lesssim 1$, but no significant evidence
for decelerated expansion. Again the results don't mean that
in the redshift ranges $0.1\lesssim z\lesssim 1$ the universe always experienced
accelerated expansion.

\begin{figure}[htbp]
\centering
\includegraphics[width=0.6\linewidth]{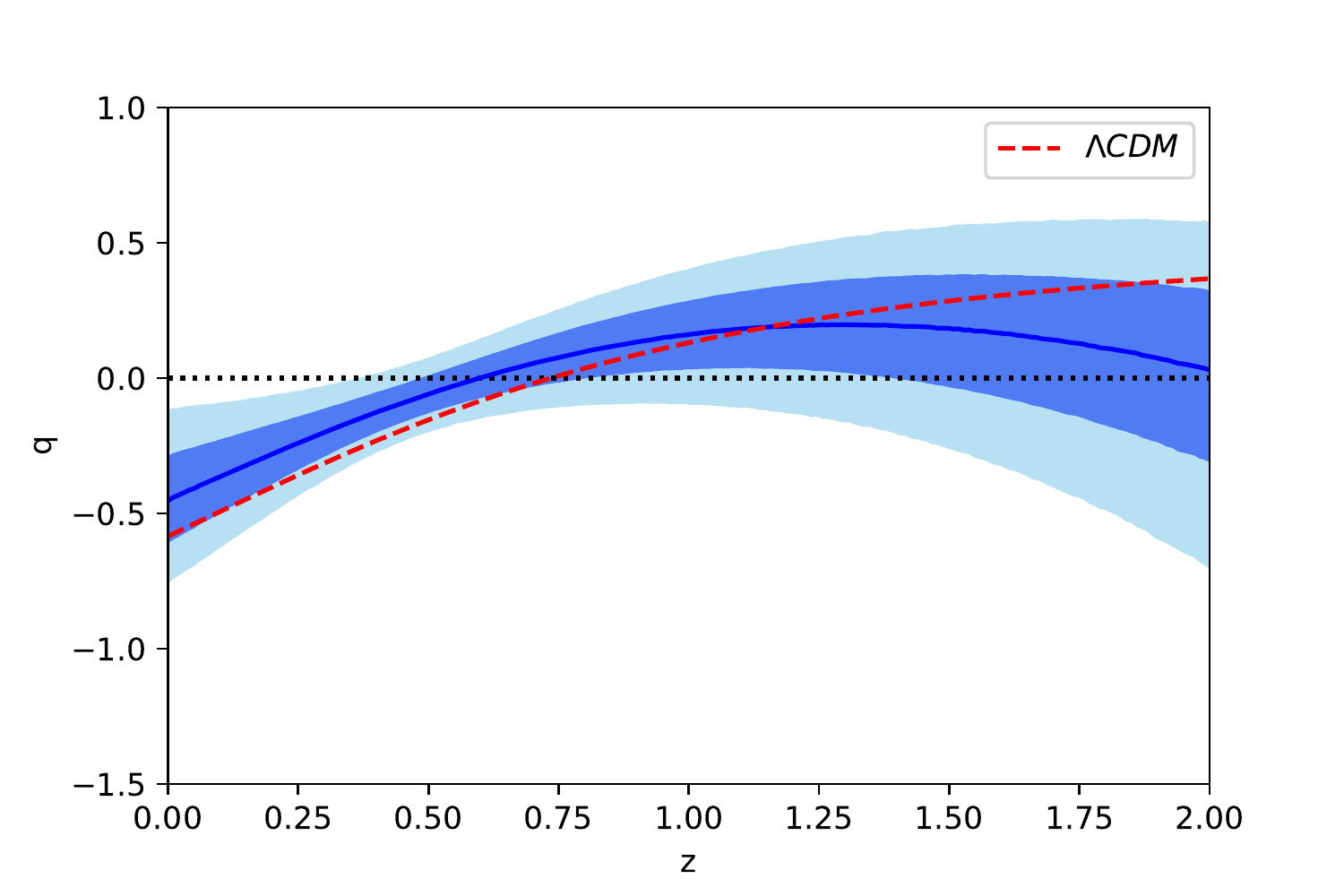}
\caption{GP reconstruction of the deceleration parameter $q(z)$ from
CCH+Pantheon+MCT. The blue solid line is the mean of the reconstruction
and the shaded areas are $1\sigma$ and $2\sigma$ errors.
The dashed line corresponds to $\Lambda CDM$ model
and the dotted line corresponds to $q(z)=0$.}
\label{gpqzfig}
\end{figure}

In order to get detailed information about
the acceleration and the transition redshift,
we reconstruct the deceleration parameter $q(z)$ from the reconstructed $E(z)$
and $E'(z)$ by using the relation $q(z)=E'(z)(1+z)/E(z)-1$,
and the result is shown in figure \ref{gpqzfig}.
We see that accelerated expansion happened
until $z\lesssim 0.3$ at the $2\sigma$ level. The mean of reconstruction suggests that
the transition from deceleration to acceleration happened at $z_t=0.60_{-0.12}^{+0.21}$. It is consistent with the result
$0.33<z_t<1.0$ obtained in \cite{Yu:2017iju} and our constraint
is more stringent due to the addition of SNe Ia data.

\section{Observational constraints on acceleration and dark energy}
\label{sec4}
To better understand the properties of acceleration and dark energy
and to measure cosmological parameters,
we use two particular parameterizations to fit the combination
of SNe Ia, $H(z)$ and BAO data in this section. We first consider a simple parametrization of the deceleration parameter $q(z)$ \cite{gong:2006tx}
\begin{equation}
\label{qzeq8}
  q(z)=\frac{1}{2}+\frac{q_1z+q_2}{(1+z)^2}.
\end{equation}
In this model, $q_0=1/2+q_2$.
Substituting equation \eqref{qzeq8} into equation \eqref{hzqzeq}, we get
\begin{equation}
\label{hzeq8}
  H(z)=H_0(1+z)^{\frac{3}{2}}\exp\left[\frac{q_2}{2}+\frac{q_1z^2-q_2}{2(1+z)^2} \right].
\end{equation}
To fit the model to CCH data, we calculate
\begin{equation}
  \chi_H^2=\sum_i\frac{[H_{obs}(z_i)-H_{th}(z_i)]^2}{\sigma_i^2}.
\end{equation}
For the SNe Ia data, we consider the distance modulus measurements from Pantheon
compilation and the expansion rate measurements from Pantheon+MCT separately.
For the Pantheon+MCT data, we calculate
\begin{equation}
  \chi_{SN}^2=\sum_{ij}[E_{obs}(z_i)-E_{th}(z_i)]C_E^{-1}(z_i,z_j)[E_{obs}(z_j)-E_{th}(z_j)],
\end{equation}
For the distance modulus data, we combine equations \eqref{hzeq8} and \eqref{dlzeq1}
to get the distance modulus $\mu_{th}=5\log_{10}[d_L(z)/\text{Mpc}]+25$,
then we calculate
\begin{equation}
  \chi_{SN}^2=\Delta\mu^T\cdot \Sigma_{\mu}^{-1}\cdot\Delta\mu,
\end{equation}
where $\Delta\mu=\mu_{obs}-\mu_{th}$ and $\Sigma_{\mu}$ is the total covariance matrix. Finally, the total chi-square is given by
\begin{equation}
  \chi^2=\chi_H^2+\chi_{SN}^2
\end{equation}
Unlike the parametrization of the equation of state, the parameters $\Omega_{m0}$ and $\Omega_b$ are absent in equation \eqref{hzeq8} and they are not model parameters,
but we need them to calculate the sound horizon at the drag redshift for the BAO parameters, so we don't use the BAO data for this model.

\begin{figure}[htbp]
\centering
\includegraphics[width=0.8\linewidth]{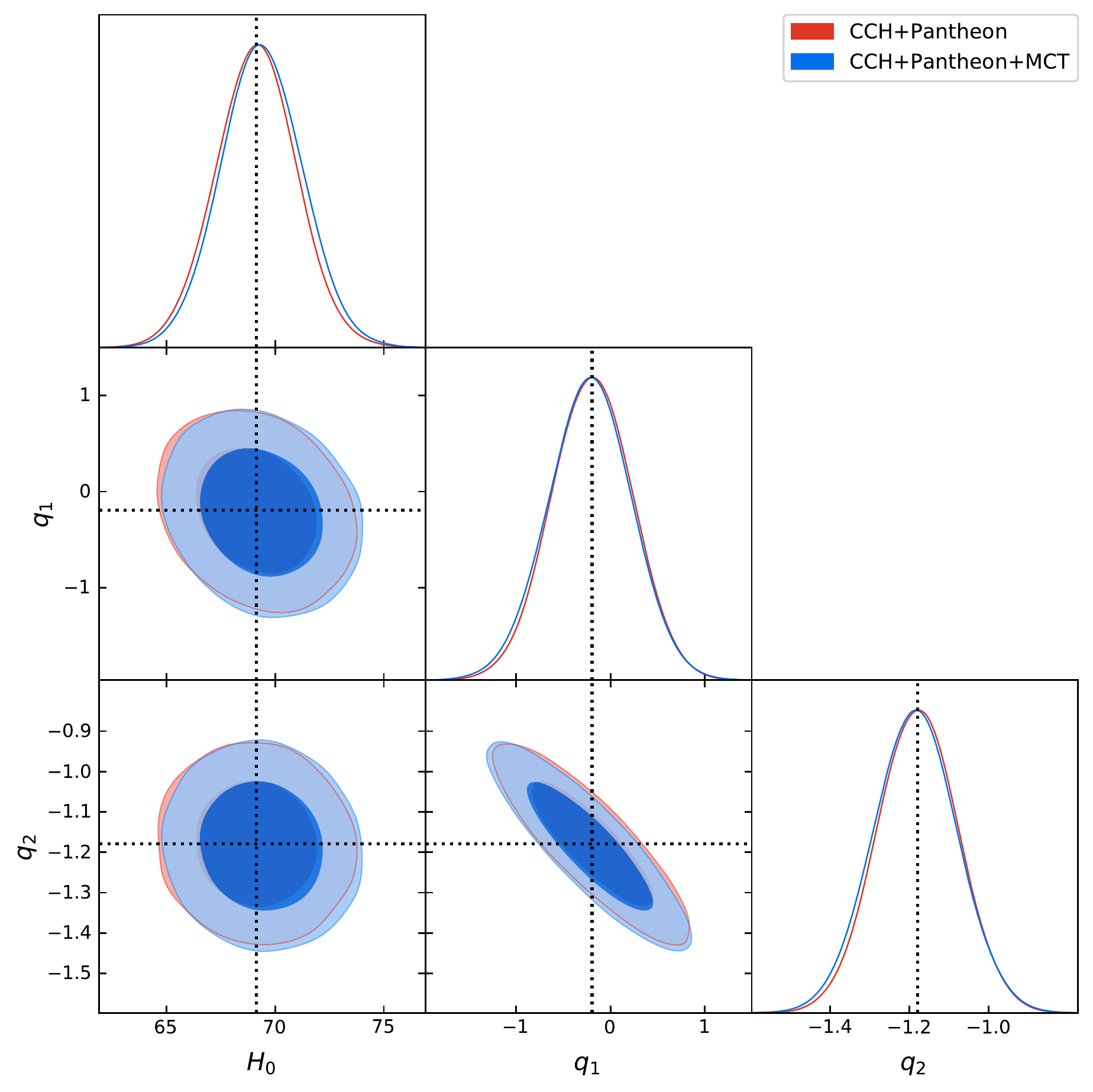}
\caption{The $1\sigma$ and $2\sigma$ contour plots for the simple $q(z)$ model.}
\label{fig5}
\end{figure}

Fitting the model to CCH+Pantheon
and CCH+Pantheon+MCT, we obtain the constraints
on the model parameters $q_1$ and $q_2$ along with the Hubble
constant and the results are shown in figure \ref{fig5}.
The $1 \sigma$ constraints on the model parameters are shown in table \ref{tab5}.
From table \ref{tab5} and figure \ref{fig5}, we see that
the constraints on the model parameters $H_0$, $q_1$ and $q_2$ are very similar with
either CCH+Pantheon or CCH+Pantheon+MCT data and the results are consistent,
so we can replace the full distance modulus data by the compressed expansion rate data.
The Hubble constant is consistent
with the Planck 2018 result.

For comparison, we also fit the curved $\Lambda$CDM model
and $Rh=ct$ model to the CCH+Pantheon data. Because the
Patheon+MCT data assume a spatially flat universe, so we don't fit
the curved $\Lambda$CDM model and $Rh=ct$ model to this data.
For the curved $\Lambda$CDM model, we get $H_0=69.4\pm 2.0$ km/s/Mpc,
$\Omega_{m0}=0.33\pm 0.06$,
$\Omega_{k0}=-0.08\pm0.16$ and $\chi^2=1050.37$,
this result is also shown in table \ref{tab6}.
For the $Rh=ct$ model, we get $H_0=62.34\pm 1.43$ km/s/Mpc and $\chi^2=1140.65$.
Both the simple $q(z)$ and the curved $\Lambda$CDM models fit the data well
and the Hubble constant is consistent with Planck 2018 result.
In terms of Akaike Information Criterion (AIC) which is
defined as $\chi^2+2n$, where $n$ is the number of parameters in the model,
we get AIC=1056.37 for the curved $\Lambda$CDM model
and AIC=1142.65 for the $Rh=ct$ model.
So comparing with the simple $q(z)$ model and the curved $\Lambda$CDM model,
the $Rh=ct$ model is strongly disfavored by the CCH+Pantheon data.

\begin{table}[htbp]
\centering
\caption{The $1\sigma$ constraints on the model parameters for the simple
$q(z)$ model. QDa denotes the data sets CCH+Pantheon and QDb denotes the data
sets CCH+Pantheon+MCT. }
\begin{tabular}{c|ccccc}
  \hline\hline
   Data sets & $H_0$ (km/s/Mpc) & $q_1$ & $q_2$ & $\chi^2$ & AIC \\
  \hline
  QDa & $69.14\pm1.86$ & $-0.19\pm0.43$ & $-1.18\pm0.10$& 1050.77 & 1056.77 \\
  QDb & $69.38\pm1.87$ & $-0.21\pm0.44$ & $-1.18\pm0.11$ & 20.86 & 26.86 \\
  \hline\hline
\end{tabular}
\label{tab5}
\end{table}

\begin{table}[htbp]
\centering
\caption{The $1\sigma$ constraints on the model parameters for the $\Lambda$CDM model. QDa denotes the data sets CCH+Pantheon and SDa denotes the data
sets CCH+BAO+Pantheon. }
\begin{tabular}{c|ccccc}
  \hline\hline
   Data sets & $H_0$ (km/s/Mpc) & $\Omega_{m0}$ & $\Omega_{k0}$ & $\chi^2$ & AIC \\
  \hline
  QDa & $69.4\pm2.0$ & $ 0.33\pm0.06$ & $-0.08\pm0.16$& 1050.37 & 1056.67 \\
  SDa & $68.9\pm1.8$ & $ 0.325\pm0.014$ & $-0.09\pm0.05$ & 1065.63 & 1071.63 \\
  \hline\hline
\end{tabular}
\label{tab6}
\end{table}

\begin{figure}[htbp]
\centering
\includegraphics[width=0.6\linewidth]{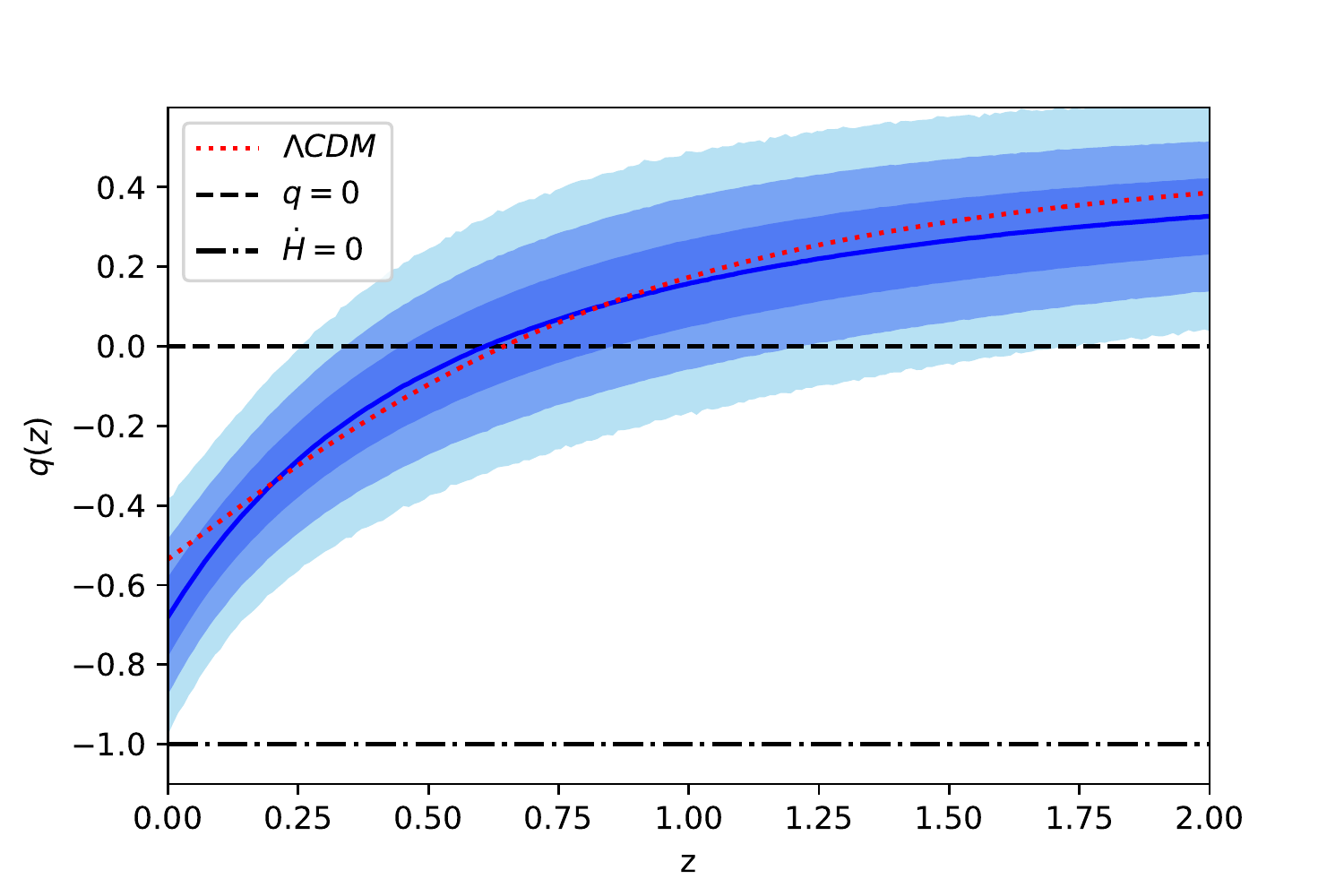}
\caption{The reconstruction of the deceleration parameter by
using the constraints from CCH+Pantheon data for the simple $q(z)$ model.
The solid line is drawn by using the best fit parameters.
The shaded areas are the $1\sigma$, $2\sigma$ and $3\sigma$ uncertainties.
The red dotted line denotes the best fit $\Lambda$CDM model.}
\label{fig10}
\end{figure}

By using the observational constraints, we reconstruct $q(z)$
and the result is shown in figure \ref{fig10}.
Figure \ref{fig10} shows $3 \sigma$ evidence
for cosmic acceleration in the redshift ranages $0 \leq z \leq0.25$
and $2 \sigma$ evidence for cosmic deceleration in the past with $z\gtrsim 1$.
The transition redshift when the Universe underwent the transition from deceleration to acceleration is $z_t=0.61_{-0.16}^{+0.24}$ at the $1 \sigma$ confidence level.
This result is consistent with that in the last section obtained from GP
reconstruction and $\Lambda$CDM model.

\begin{figure}[htbp]
\centering
\includegraphics[width=0.8\linewidth]{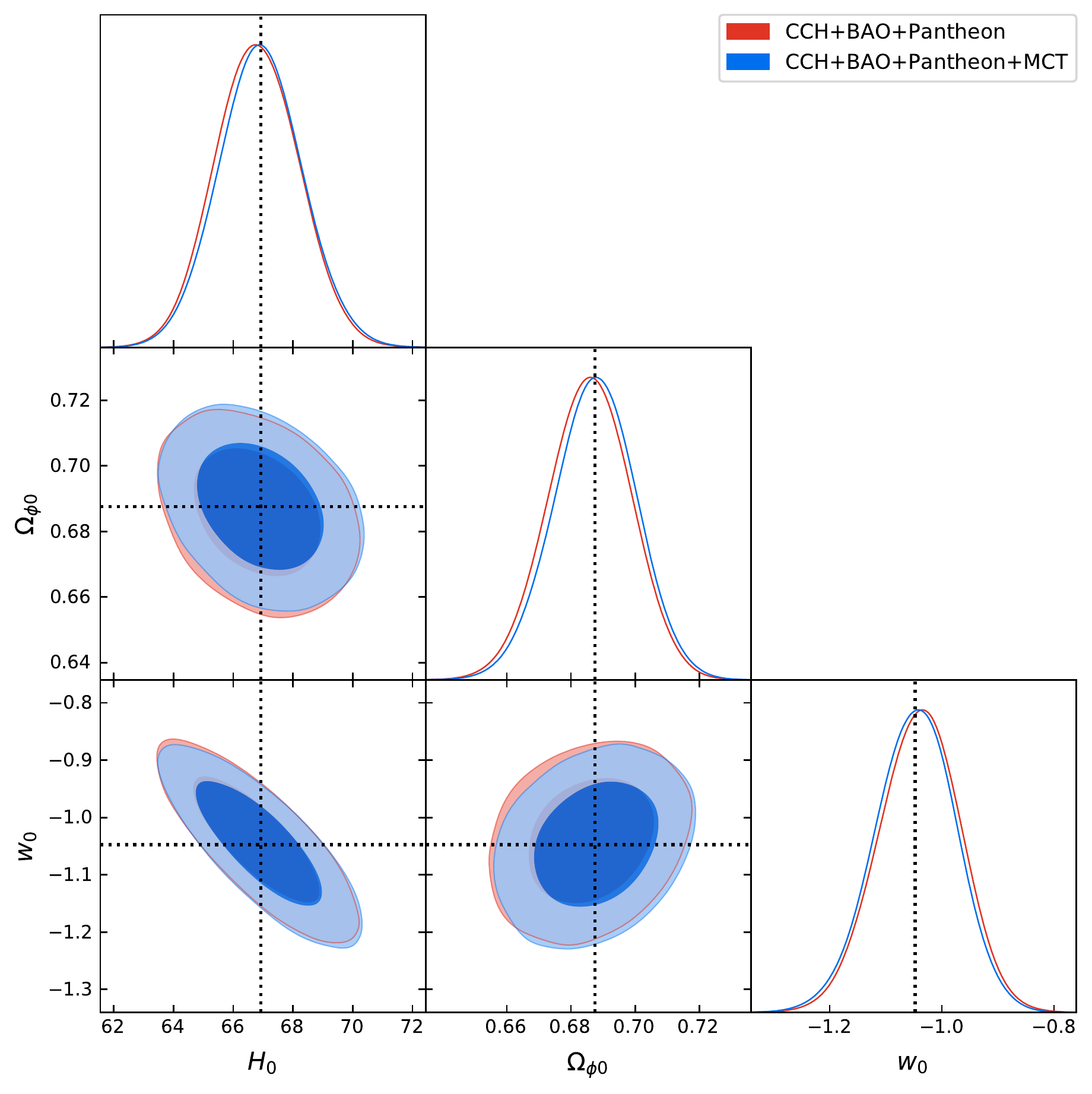}
\caption{The $1\sigma$ and $2\sigma$ contour plots for the SSLCPL model.}
\label{fig11}
\end{figure}

To use the BAO data and measure the cosmological parameters,
now we consider the SSLCPL model \cite{Gao:2012ef, Gong:2013bn}.
This model approximates the dynamics of general
thawing scalar fields over a large redshift range
and it has only one parameter for $w(z)$. It has the same form as the
commonly used Chevallier-Polarski-Linder (CPL) model which parameterizes
the equation of state parameter as \cite{Chevallier:2000qy,Linder:2002et}
\begin{equation}
\label{cpleq1}
w(z)=w_0+w_a\frac{z}{1+z}.
\end{equation}
For the SSLCPL model, the parameter $w_a$ is not an independent parameter, i.e.,
\begin{equation}
  w_a=6(1+w_0)\frac{(\Omega_{\phi0}^{-1}-1)[\sqrt{\Omega_{\phi0}}-\tanh^{-1}(\sqrt{\Omega_{\phi0}})]}
  {\Omega_{\phi0}^{-1/2}-(\Omega_{\phi0}^{-1}-1)\tanh^{-1}(\sqrt{\Omega_{\phi0}})},
\end{equation}
where $\Omega_{\phi0}$ is the dark energy density
normalized by the current critical energy density.
The SSLCPL model has only one free parameter $w_0$
and it reduces to the $\Lambda$CDM model when the parameter $w_0=-1$. With this explicit degeneracy relation between $w_0$ and $w_a$, we expect to
get tighter constraints on $\Omega_{\phi0}$ and $w_0$ for the SSLCPL model
than the CPL model does. For a spatially flat universe, the
Friedmann equation for the CPL parametrization becomes
\begin{equation}
\begin{split}
   \frac{H^2}{H_0^2}= & \Omega_{r0} (1+z)^4+\Omega_{m0}(1+z)^3+ \\
     &(1-\Omega_{r0}-\Omega_{m0})(1+z)^{3(1+w_0+w_a)}\exp(\frac{-3w_az}{1+z}),
\end{split}
\end{equation}
where $\Omega_{r0}$ is radiation density parameter,
$\Omega_{m0}$ is matter density parameter and
$1-\Omega_{r0}-\Omega_{m0}=\Omega_{\phi0}$ is the dark energy density parameter.
To fit the flat SSLCPL model to the observational data and obtain the best
fit parameters, we minimize
\begin{equation}
  \chi^2=\chi_H^2+\chi_{SN}^2+\chi_{BAO}^2.
\end{equation}
The results for fitting both the CCH+BAO+Pantheon data (we label this
data sets as SDa) and CCH+BAO+Pantheon+MCT data
(we label this data sets as SDb)
are shown in figure \ref{fig11} and table \ref{tab11}.
From figure \ref{fig11} and table \ref{tab11},
we see that the constraints on the model parameters from both SDa and SDb data
are very similar and they are consistent, and both results are consistent
with flat $\Lambda$CDM model. As shown in tables \ref{tab6} and \ref{tab11},
both flat SSLCPL and curved $\Lambda$CDM model fit the observational data well.

\begin{table}[htbp]
  \centering
  \caption{The $1\sigma$ constraints on the model parameters for the SSLCPL model.
  SDa denotes the data sets CCH+BAO+Pantheon and SDb denotes the data sets CCH+BAO+Pantheon+MCT. }
  \begin{tabular}{c|ccccc}
    \hline\hline
     Data sets & $H_0$ (km/s/Mpc) & $\Omega_{\phi0}$ & $w_0$ & $\chi^2$ &AIC \\
    \hline
    SDa & $66.79\pm 1.40$ & $0.687\pm0.013$ & $-1.03\pm0.07$ &1068.48 & 1074.48 \\
    \hline
    SDb & $66.90\pm1.39$ & $0.688\pm0.013$ & $-1.04\pm0.07$ &38.85 & 44.85\\
    \hline
  \end{tabular}
\label{tab11}
\end{table}

\section{Discussion}
\label{sec5}

The null hypothesis that cosmic acceleration never happened gives the kinematic
bound $E(z)\ge 1+z$. In standard cosmology, the null hypothesis $q(z)\ge 0$ is
equivalent to the strong energy condition $\rho+3p \geq 0$. The six $E(z)$
data from Patheon+MCT can be compared directly with the lower bound to
give direct evidence of cosmic acceleration. The five low redshfit
data points with $z<1$ lie outside the lower bound at the $3\sigma$ confidence level,
and the only one high redshift data crosses the bound at the $1\sigma$ level.
Therefore, we have $3\sigma$ direct evidence of cosmic acceleration.
This direct evidence does not assume any gravitational
theory or cosmological model. The only caveat from this direct evidence
is that the $E(z)$ assumes a spatially flat universe. Although
there is no strong evidence for decelerated expansion, it does
not mean the cosmic acceleration started at least from $z\sim 0.9$
or there is no decelerated expansion in the redshift ranges $0<z<0.9$
because of the integration effect of the deceleration parameter.
Due to the integration effect,
even if the transition from cosmic acceleration to deceleration happened
at the redshift $z\sim 0.6$, the expansion rate remains outside the bound
until $z\gtrsim 2$.
The direct evidence does exclude the $Rh=ct$ at the $3\sigma$ confidence level.
Comparing the $Rh=ct$ model with $\Lambda$CDM model by the $\chi^2$ statistics,
the $Rh=ct$ model is also strongly disfavoured by the $E(z)$ data.
We also use the CCH data to give direct evidence of cosmic acceleration.
Due to large error bars in the data and the uncertainties in the value of the Hubble
constant, only several data points lie outside the lower bound.
Those data points may be outliers, so the evidence from CCH data is
not convincing.

The GP method was used to reconstruct the $E(z)$ and $q(z)$ functions from
the CCH and Pantheon+MCT data. The Hubble constant $H_0=67.46\pm 4.75$ km/s/Mpc
inferred from the reconstructed $H(z)$ by CCH data is consistent
with the Planck 2018 result, but it has a little tension with the local measurement
even though the error bar is big. The reconstructed $E(z)$ shows
more than $3\sigma$ direct evidence for cosmic acceleration up
to the redshift $z\sim 1$, and the reconstructed $q(z)$ function
gives the transition redshift $z_t=0.60_{-0.12}^{+0.21}$
at which the expansion of the Universe
underwent the transition from acceleration to deceleration.

Fitting the simple two-parameter parametrization $q(z)=1/2+(q_1z+q_2)/(1+z)^2$
to CCH+Patheon and CCH+Patheon+MCT data we get consistent constraints
on the model parameters. The best fit Hubble
constant is $H_0=69.14\pm 1.86$ km/s/Mpc. This
value is consistent with the Planck 2018 result and has a little tension
with the local measurement.
By using the fitted parameters from CCH+Patheon, we reconstruct $q(z)$
and get the transition redshift $z_t=0.61^{+0.24}_{-0.16}$ which
is consistent with that from GP method.
We also fit the SSLCPL model to the combination of CCH, BAO and SNe Ia data,
and we get consistent results with either Pantheon or Patheon+MCT data.
The Hubble constant $H_0=66.79\pm 1.4$ km/s/Mpc from fitting
the SSLCPL model to the CCH+BAO+Pantheon data
is consistent with the Planck 2018 result and is in tension
with the local measurement at $3.6\sigma$ confidence level.
The addition of the type Ia SNe data helps the CCH and BAO
data to tighten the error bar on the Hubble constant,
but it does not affect the value of the Hubble constant
because of the arbitrary normalization of the luminosity distance.

In conclusion, the expansion rate measured from Pantheon+MCT gives
more than $3\sigma$ direct evidence for cosmic acceleration. In fitting
cosmological models, we can use the six compressed data points on the expansion rate
instead of the full Pantheon compilation. The CCH and BAO data prefers
lower value of the Hubble constant which is consistent with Planck 2018 result.

\acknowledgments
This research was supported in part by the National Natural Science
Foundation of China under Grant No. 11875136 and
the Major Program of the National Natural Science Foundation of China under Grant No. 11690021.

\appendix

\section{Observational data}
\label{appdxa}

The Hubble parameter directly probes the expansion history of the Universe by its definition $H=\dot{a}/a$, where $a$ denotes the cosmic scale factor and $\dot{a}$ is its rate of change with respect to the cosmic time $t$.
Since the Hubble parameter is related with the differential redshift time as
\begin{equation}
  H(z)=-\frac{1}{(1+z)}\frac{dz}{dt}\approx-\frac{1}{(1+z)}\frac{\Delta z}{\Delta t},
\end{equation}
and $dz$ is obtained from spectroscopic surveys, so a measurement
of $dt$ gives the Hubble parameter which is independent of the cosmological model.
Based on the spectroscopic differential evolution of passively evolving
galaxies, CCH method obtains the expansion rate $dz/dt$ by taking a pair
of massive and passively evolving galaxies at two different redshifts \cite{Jimenez:2001gg}.
We show the 31 CCH data points of $H(z)$ compiled by \cite{Farooq:2016zwm,Yu:2017iju,Marra:2017pst,Haridasu:2018gqm,Gomez-Valent:2018hwc,Gomez-Valent:2018gvm}
in table \ref{hztable}. These data cover a redshift range up to $z\sim 2$ and are obtained without assuming any particular cosmological model. There exit
systematic uncertainties associated with the stellar population synthesis models like
BC03 \cite{Bruzual:2003tq} and MaStro \cite{Maraston:2011sq}, and a possible contamination due to young underlying stellar
components in quiescent galaxies \cite{Gomez-Valent:2018gvm}, so the data
is model dependent in this sense. Here we use the measurements on $H(z)$
with the BC03 model. To keep the data to be model independent as minimum
as possible, we don't use the $H(z)$ data determined by BAO measurements in this paper.

\begin{table}[htbp]
\centering
\caption{The 31 CCH data with the BC03 model. The unit for $H(z)$ is km/s/Mpc.}
\begin{tabular}{ccccccccc}
\hline\hline
$z$ & $H(z)$ & $\sigma_{H(z)}$& Ref. & &$z$ & $H(z)$ & $\sigma_{H(z)}$& Ref. \\
\hline
 0.07 & 69.0 & 19.6 & \cite{Zhang:2012mp} & &0.4783 & 80.9 & 9.0  & \cite{Moresco:2016mzx}\\
 0.09 & 69.0 & 12.0 & \cite{Simon:2004tf} & & 0.48 & 97.0 & 62.0 & \cite{Stern:2009ep}  \\
 0.12 & 68.6 & 26.2 & \cite{Zhang:2012mp}  & & 0.593 & 104.0 & 13.0 & \cite{Moresco:2012jh} \\
 0.17 & 83.0 & 8.0 & \cite{Simon:2004tf}  & & 0.68 & 92.0 & 8.0 & \cite{Moresco:2012jh} \\
 0.179 & 75.0 & 4.0 & \cite{Moresco:2012jh}  & & 0.781 & 105.0 & 12.0  & \cite{Moresco:2012jh} \\
 0.199 & 75.0 & 5.0 & \cite{Moresco:2012jh}  & & 0.875 & 125.0 & 17.0 & \cite{Moresco:2012jh} \\
 0.2 & 72.9 & 29.6 & \cite{Zhang:2012mp}  & & 0.88 & 90.0 & 40.0 & \cite{Stern:2009ep} \\
 0.27 & 77.0 & 14.0 & \cite{Simon:2004tf}  & & 0.9 & 117.0 & 23.0 & \cite{Simon:2004tf} \\
 0.28 & 88.8 & 36.6 & \cite{Zhang:2012mp}  & & 1.037 & 154.0 & 20.0 & \cite{Moresco:2012jh} \\
 0.352 & 83.0 & 14.0 & \cite{Moresco:2012jh}  & & 1.3 & 168.0 & 17.0 & \cite{Simon:2004tf} \\
 0.3802 & 83.0 & 13.5 & \cite{Moresco:2016mzx} & & 1.363 & 160.0 & 33.6 & \cite{Moresco:2015cya} \\
 0.4 & 95.0 & 17.0 & \cite{Simon:2004tf}  & & 1.43 & 177.0 & 18.0 & \cite{Simon:2004tf} \\
 0.4004 & 77.0 & 10.2 & \cite{Moresco:2016mzx}  & & 1.53 & 140.0 & 14.0 & \cite{Simon:2004tf} \\
 0.4247 & 87.1 & 11.2  & \cite{Moresco:2016mzx} & & 1.75 & 202.0 & 40.0 & \cite{Simon:2004tf} \\
 0.4497 & 92.8 & 12.9  & \cite{Moresco:2016mzx} & & 1.965 & 186.5 & 50.4  & \cite{Moresco:2015cya} \\
 0.47 & 89.0 & 49.6  & \cite{Ratsimbazafy:2017vga} & & & & & \\
 \hline\hline
\end{tabular}
\label{hztable}
\end{table}

The Pantheon sample \cite{Scolnic:2017caz} is the largest SNe Ia sample
which includes 1048 spectroscopically confirmed SNe Ia and the furthest SN reaches approximately redshift $z\sim 2.3$.
It consists of 279 spectroscopically confirmed SNe Ia with
redshift $0.03<z<0.68$ discovered by the Pan-STARRS1 Medium Deep Survey \cite{Rest:2013mwz},
samples of SNe Ia from the Harvard Smithsonian Center for Astrophysics SN surveys \cite{Hicken:2009df}, the Carnegie SN Project \cite{Stritzinger:2011qd}, the Sloan digital sky survey \cite{Kessler:2009ys} and the SN legacy survey \cite{Conley:2011ku},
and high-z data with the redshift $z > 1.0$ from the Hubble space telescope cluster SN survey \cite{Suzuki:2011hu}, GOODS \cite{Riess:2006fw} and CANDELS/CLASH survey \cite{Rodney:2014twa,Graur:2013msa}. The calibration systematics is reduced substantially by cross-calibrating all of the SN samples.
The distance modulus $\mu$ of SNe Ia was derived from the observation of light curves through the SALT2 light-curve fitter
\begin{equation}
  \mu_{obs}=m_B-M_B+\alpha\cdot X_1-\beta\cdot C+\Delta_M+\Delta_B,
\end{equation}
where $m_B$ corresponds to the observed peak magnitude in rest-frame $B$ band,
$X_1$ is the time stretching of the light curve, $C$ describes the supernova color at maximum brightness, $M_B$ is the absolute $B$-band magnitude of a fiducial SN Ia with
$X_1=0$ and $C=0$, $\Delta_M$ is a distance correction based on the host-galaxy mass of the SN and $\Delta_B$ is a distance correction based on predicted biases from
simulations. The parameters $\alpha$ and $\beta$ characterize luminosity-stretch,
and luminosity-color relations.
Since the absolute magnitude of a SN Ia is degenerated with the Hubble constant, the corrected magnitudes $\mu+M_B$ are given for cosmological model fitting \cite{Scolnic:2017caz}. The nuisance parameters $\alpha$, $\beta$ and $H_0$ should be marginalized. The statistical uncertainty and systematic uncertainty are also given in ref. \cite{Scolnic:2017caz}. The total uncertainty matrix of the distance modulus is given by
\begin{equation}
  \Sigma_{\mu}=D_{stat}+C_{sys},
\end{equation}
where the statistical matrix $D_{stat}$ has only a diagonal component
and $C_{sys}$ is the systematic covariance.
We take into account all the statistical uncertainties as described by their full covariance matrix.

Recently, Riess et al. combine the Pantheon sample with 15 SNe Ia at redshift $z>1$
discovered in the CANDELS and CLASH Multi-Cycle Treasury (MCT) programs
using WFC3 on the Hubble Space Telescope and compress the raw distance measurements
to expansion rate $E(z)$ at six redshifts in the range $0.07<z<1.5$
by assuming a flat universe with $\Omega_k=0$ \cite{Riess:2017lxs}, the results and the correlation matrix of $E(z)$ are shown in table \ref{eztable}. Because of the assumption of a flat universe, the
results of $E(z)$ are cosmological model dependent in this sense.
The last point $E(z=1.5)$ is not Gaussian,
the symmetrization of the upper and lower bounds gives
$E(1.5)=2.924\pm 0.675$ \cite{Haridasu:2018gqm},
or $E(1.5)=2.67\pm 0.675$ \cite{Pinho:2018unz}, and the Gaussian approximation
is $E(1.5)=2.78\pm 0.59$ \cite{Gomez-Valent:2018gvm}.
\begin{table}[htbp]
  \centering
  \caption{Pantheon+MCT SN Ia measurements of $E(z)$ \cite{Riess:2017lxs}.}
  \begin{tabular}{cc|cccccc}
    \hline\hline
     $z$ & $E(z)$ &\multicolumn{6}{c}{Correlation Matrix} \\
    \hline
    0.07 & $0.994\pm 0.023$ & 1.00 &  &  &  &  &  \\
    0.2  & $1.113\pm 0.020$ & 0.40 & 1.00 &  &  &  & \\
    0.35 & $1.122\pm 0.037$ & 0.52 & -0.13 & 1.00 &  &  & \\
    0.55 & $1.369\pm 0.063$ & 0.35 & 0.35 & -0.18 & 1.00 &  & \\
    0.9  & $1.54\pm 0.12$ & 0.02 & -0.08 & 0.19 & -0.41 & 1.00 & \\
    1.5  & $2.69^{+0.86}_{-0.52}$ & 0.00 & -0.06 & -0.05 & 0.16 & -0.21 & 1.00\\
    \hline
  \end{tabular}
  \label{eztable}
\end{table}

BAO is a powerful standard ruler to probe the angular diameter distance and the Hubble parameter evolution. The isotropic and anisotropic BAO measurements are summarized in tables \ref{baotab1} and \ref{baotab2}, respectively \cite{Evslin:2017qdn}.
The covariance matrix associated with the data in table \ref{baotab2} is
  \[\bm{C}=
    \begin{pmatrix}
      0.0150 & -0.0357 & 0.0071 & -0.0100 & 0.0032 & -0.0036 & 0 & 0 \\
      -0.0357 & 0.5304 & -0.0160 & 0.1766 & -0.0083 & 0.0616 & 0 & 0 \\
      0.0071 & -0.0160 & 0.0182 & -0.0323 & 0.0097 & -0.0131 & 0 & 0 \\
      -0.0100 & 0.1766 & -0.0323 & 0.3267 & -0.0167 & 0.1450 & 0 & 0 \\
      0.0032 & -0.0083 & 0.0097 & -0.0167 & 0.0243 & -0.0352 & 0 & 0 \\
      -0.0036 & 0.0616 & -0.0131 & 0.1450 & -0.0352 & 0.2684 & 0 & 0 \\
      0 & 0 & 0 & 0 & 0 & 0 & 0.1358 & -0.0296 \\
      0 & 0 & 0 & 0 & 0 & 0 & -0.0296 & 0.0492 \\
    \end{pmatrix}.\]

\begin{table}[htbp]
  \centering
  \caption{Isotropic BAO data.}
  \begin{tabular}{cccc}
    \hline\hline
    Data set & Redshift & $D_V(z)/r_d$ & Ref. \\
    \hline
    6dF & z=0.106 & $2.98\pm0.13$ & \cite{Beutler:2011hx} \\
    MGS & z=0.15 & $4.47\pm0.17$ & \cite{Ross:2014qpa} \\
    eBOSS quasars & z=1.52 & $26.1\pm1.1$ & \cite{Ata:2017dya} \\
    \hline
  \end{tabular}
\label{baotab1}
\end{table}

\begin{table}[htbp]
  \centering
  \caption{Anisotropic BAO data. In the third column, $A$ means $D_{A}(z)/r_d$
  and $H$ means $D_{H}(z)/r_d$.}
  \begin{tabular}{cccc}
    \hline\hline
    Data set & Redshift & $D_{A/H}(z)/r_d$ & Ref. \\
    \hline
    BOSS DR12 & z=0.38 & $7.42 (A)$ & \cite{Alam:2016hwk} \\
    BOSS DR12 & z=0.38 & $24.97 (H)$ & \cite{Alam:2016hwk} \\
    BOSS DR12 & z=0.51 & $8.85 (A)$ & \cite{Alam:2016hwk} \\
    BOSS DR12 & z=0.51 & $22.31 (H)$ & \cite{Alam:2016hwk} \\
    BOSS DR12 & z=0.61 & $9.69 (A)$ & \cite{Alam:2016hwk} \\
    BOSS DR12 & z=0.61 & $20.49 (H)$ & \cite{Alam:2016hwk} \\
    BOSS DR12 & z=2.4 & $10.76 (A)$ & \cite{Bourboux:2017cbm} \\
    BOSS DR12 & z=2.4 & $8.94 (H)$ & \cite{Bourboux:2017cbm} \\
    \hline
  \end{tabular}
\label{baotab2}
\end{table}

In a spatially flat universe, the Hubble distance is
\begin{equation}
\label{dhzeq1}
  D_H(z)=\frac{c}{H(z)},
\end{equation}
the angular diameter distance is
\begin{equation}
\label{dazeq1}
  D_A(z)=\frac{c}{1+z}\int_0^z\frac{dx}{H(x)},
\end{equation}
the luminosity distance is
\begin{equation}
\label{dlzeq1}
  d_L(z)=c(1+z)\int_0^z\frac{dx}{H(x)},
\end{equation}
and the effective distance $D_V(z)$ is \cite{Eisenstein:2005su}
\begin{equation}
\label{dvzeq1}
  D_V(z)=\left[\frac{d_L^2(z)}{(1+z)^2}\frac{cz}{H(z)}\right]^{1/3}.
\end{equation}
The sound horizon at the drag redshift $z_d$ is
\begin{equation}
\label{rdeq1}
  r_d=\frac{c}{\sqrt{3}}\int_{z_d}^{\infty}\frac{dz}{\sqrt{1+(3\Omega_b/4\Omega_\gamma)/(1+z)}H(z)},
\end{equation}
and the drag redshift $z_d$ is fitted as \cite{Eisenstein:1997ik}
\begin{equation}
\label{zdeq1}
  z_d=\frac{1291(\Omega_mh^2)^{0.251}}{1+0.659(\Omega_mh^2)^{0.828}}[1+b_1(\Omega_bh^2)^{b_2}],
\end{equation}
where
\begin{equation}
  b_1=0.313(\Omega_bh^2)^{-0.419}[1+0.607(\Omega_mh^2)^{0.674}],
\end{equation}
and
\begin{equation}
  b_2=0.238(\Omega_mh^2)^{0.223}.
\end{equation}
In this work we take $\Omega_bh^2=0.02236$ and $\Omega_\gamma h^2=2.469\times10^{-5}$ \cite{Aghanim:2018eyx}, where the dimensionless parameter $h=H_0/(100$ km/s/Mpc).
To use the data, we calculate
\begin{equation}
  \chi_{BAO}^2=\chi_{iso}^2+\chi_{aniso}^2,
\end{equation}
\begin{equation}
  \chi_{iso}^2=\sum_{i}(\frac{v_i-d_i^{iso}}{\sigma_i})^2,
\end{equation}
\begin{equation}
  \chi_{aniso}^2=(\bm{w}-\bm{d}^{aniso})^T\bm{C}^{-1}(\bm{w}-\bm{d}^{aniso}),
\end{equation}
where the vectors $\bm{d}^{iso}$ and $\bm{d}^{aniso}$ are the isotropic and anisotropic data from tables \ref{baotab1} and \ref{baotab2}, respectively, and $\bm{v}$ and $\bm{w}$ are the predictions for these vectors in a given cosmological model.

\section{Gaussian process method}
\label{appdxb}

Because of insufficient and low quality of observational data,
we use the GP method to find a smooth function $f(x)$
that best represents a set of observational data points $f(x_i)\pm\sigma_i$.
The GP method assume that the value of the function $f(x)$ at any point $x$
follows a Gaussian distribution. At each $z_i$, the value of $f(z_i)$
is drawn from a Gaussian distribution with mean $u(z_i)$ and
variance $k(z_i,z_i)$. Besides, $f(z_i)$ and $f(z_j)$ are
correlated by the covariance function (or kernel function) $k(z_i,z_j)$.

A GP is written as
\begin{equation}
  f(x)\sim \text{GP}(\mu(x),k(x,x')),
\end{equation}
so the kernel function plays a crucial role in the GP method and must be selected beforehand. In this sense, GP is model dependent although it is independent of
cosmological models. There are three widely used kernel functions with two degrees
of freedom. The Gaussian/squared-exponential kernel
\begin{equation}
\label{gskernel}
k(x_i,x_j)=\sigma_f^2\,\exp\left(-\frac{(x_i-x_j)^2}{2l_f^2}\right),
\end{equation}
where $\sigma_f$ and $l_f$ are hyperparameters. The Cauchy kernel
\begin{equation}
\label{cckernel}
k(x_i,x_j)=\frac{\sigma_f^2 l_f}{(x_i-x_j)^2+l_f^2}.
\end{equation}
The Mat\'{e}rn kernel
\begin{equation}
\label{mtkernel}
k(x_i,x_j)=\sigma_f^2\frac{2^{1-\nu}}{\Gamma(\nu)}\left(\frac{\sqrt{2\nu(x_i-x_j)^2}}{l_f}\right)^\nu\,
K_\nu \left(\frac{\sqrt{2\nu (x_i-x_j)^2}}{l_f}\right),
\end{equation}
where $K_\nu$ is the modified Bessel function with $\nu$ being positive.
Here we choose the Gaussian kernel.
The hyperparameters are determined from the observed data
by minimizing the log likelihood function
\begin{equation}
  \begin{split}
   \ln\mathcal{L}=&\ln p(\bm{y}|\bm{X},\sigma_f,l) \\
         =&-\frac{1}{2}(\bm{y}-\bm{\mu})^T[K(\bm{X,X})+C]^{-1}(\bm{y}-\bm{\mu})\\
          &-\frac{1}{2}\ln|K(\bm{X,X})+C|-\frac{n}{2}\ln(2\pi),
  \end{split}
\end{equation}
where $\bm{X}=[x_1,x_2,...,x_n]^T$ are the inputs, $K(\bm{X,X})$ is the covariance matrix with components $k(x_i,x_j)$, $\bm{y}$ is the vector of observed data and $C$ is the covariance matrix of the observed data.

To predict the function values $\bm{f_*}=[f_{*1},f_{*2},...,f_{*m}]^T$
at the test locations $\bm{X_*}=[x_{n+1},x_{n+2},...,x_{n+m}]^T$,
the predictive normal distribution is
\begin{equation}
  p(\bm{f_*}|\bm{X},\bm{y},\bm{X_*})=\mathcal{N}(\hat{\bm{\mu}},\hat{\Sigma}),
\end{equation}
\begin{equation}
  \hat{\bm{\mu}}=K(\bm{X_*},\bm{X})^T(K(\bm{X},\bm{x}+C)^{-1}(\bm{y}-\bm{\mu(X)})+\bm{\mu}(\bm{X_*}),
\end{equation}
\begin{equation}
  \hat{\Sigma}=K(\bm{X_*,X_*})-K(\bm{X_*},\bm{X})^T(K(\bm{X,X})+C)^{-1}K(\bm{X_*,X}).
\end{equation}

The public available python package GaPP \cite{Seikel:2012uu} is used to do the GP reconstruction in section \ref{sec3}.


\begin{thebibliography}{100}

\bibitem{Riess:1998cb}
{\scshape Supernova Search Team} collaboration, \emph{{Observational evidence
  from supernovae for an accelerating universe and a cosmological constant}},
  \href{https://doi.org/10.1086/300499}{\emph{Astron. J.} {\bfseries 116}
  (1998) 1009} [\href{https://arxiv.org/abs/astro-ph/9805201}{{\ttfamily
  astro-ph/9805201}}].

\bibitem{Perlmutter:1998np}
{\scshape Supernova Cosmology Project} collaboration, \emph{{Measurements of
  $\Omega$ and $\Lambda$ from 42 high redshift supernovae}},
  \href{https://doi.org/10.1086/307221}{\emph{Astrophys. J.} {\bfseries 517}
  (1999) 565} [\href{https://arxiv.org/abs/astro-ph/9812133}{{\ttfamily
  astro-ph/9812133}}].

\bibitem{Dvali:2000hr}
G.~Dvali, G.~Gabadadze and M.~Porrati, \emph{{4-D gravity on a brane in 5-D
  Minkowski space}},
  \href{https://doi.org/10.1016/S0370-2693(00)00669-9}{\emph{Phys. Lett. B.}
  {\bfseries 485} (2000) 208}
  [\href{https://arxiv.org/abs/hep-th/0005016}{{\ttfamily hep-th/0005016}}].

\bibitem{Carroll:2003wy}
S.~M. Carroll, V.~Duvvuri, M.~Trodden and M.~S. Turner, \emph{{Is cosmic speed
  - up due to new gravitational physics?}},
  \href{https://doi.org/10.1103/PhysRevD.70.043528}{\emph{Phys. Rev. D}
  {\bfseries 70} (2004) 043528}
  [\href{https://arxiv.org/abs/astro-ph/0306438}{{\ttfamily
  astro-ph/0306438}}].

\bibitem{Nojiri:2003ft}
S.~Nojiri and S.~D. Odintsov, \emph{{Modified gravity with negative and
  positive powers of the curvature: Unification of the inflation and of the
  cosmic acceleration}},
  \href{https://doi.org/10.1103/PhysRevD.68.123512}{\emph{Phys. Rev.}
  {\bfseries D68} (2003) 123512}
  [\href{https://arxiv.org/abs/hep-th/0307288}{{\ttfamily hep-th/0307288}}].

\bibitem{Starobinsky:2007hu}
A.~A. Starobinsky, \emph{{Disappearing cosmological constant in f(R) gravity}},
  \href{https://doi.org/10.1134/S0021364007150027}{\emph{JETP Lett.} {\bfseries
  86} (2007) 157} [\href{https://arxiv.org/abs/0706.2041}{{\ttfamily
  0706.2041}}].

\bibitem{Hu:2007nk}
W.~Hu and I.~Sawicki, \emph{{Models of f(R) Cosmic Acceleration that Evade
  Solar-System Tests}},
  \href{https://doi.org/10.1103/PhysRevD.76.064004}{\emph{Phys. Rev. D}
  {\bfseries 76} (2007) 064004}
  [\href{https://arxiv.org/abs/0705.1158}{{\ttfamily 0705.1158}}].

\bibitem{deRham:2010kj}
C.~de~Rham, G.~Gabadadze and A.~J. Tolley, \emph{{Resummation of Massive
  Gravity}}, \href{https://doi.org/10.1103/PhysRevLett.106.231101}{\emph{Phys.
  Rev. Lett.} {\bfseries 106} (2011) 231101}
  [\href{https://arxiv.org/abs/1011.1232}{{\ttfamily 1011.1232}}].

\bibitem{Gong:2012yv}
Y.~Gong, \emph{{Cosmology in massive gravity}},
  \href{https://doi.org/10.1088/0253-6102/59/3/13}{\emph{Commun. Theor. Phys.}
  {\bfseries 59} (2013) 319} [\href{https://arxiv.org/abs/1207.2726}{{\ttfamily
  1207.2726}}].

\bibitem{Weinberg:1988cp}
S.~Weinberg, \emph{{The Cosmological Constant Problem}},
  \href{https://doi.org/10.1103/RevModPhys.61.1}{\emph{Rev. Mod. Phys.}
  {\bfseries 61} (1989) 1}.

\bibitem{Ratra:1987rm}
B.~Ratra and P.~Peebles, \emph{{Cosmological Consequences of a Rolling
  Homogeneous Scalar Field}},
  \href{https://doi.org/10.1103/PhysRevD.37.3406}{\emph{Phys. Rev. D}
  {\bfseries 37} (1988) 3406}.

\bibitem{Wetterich:1987fm}
C.~Wetterich, \emph{{Cosmology and the Fate of Dilatation Symmetry}},
  \href{https://doi.org/10.1016/0550-3213(88)90193-9}{\emph{Nucl. Phys. B}
  {\bfseries 302} (1988) 668}.

\bibitem{Caldwell:1997ii}
R.~Caldwell, R.~Dave and P.~J. Steinhardt, \emph{{Cosmological imprint of an
  energy component with general equation of state}},
  \href{https://doi.org/10.1103/PhysRevLett.80.1582}{\emph{Phys. Rev. Lett.}
  {\bfseries 80} (1998) 1582}
  [\href{https://arxiv.org/abs/astro-ph/9708069}{{\ttfamily
  astro-ph/9708069}}].

\bibitem{Zlatev:1998tr}
I.~Zlatev, L.-M. Wang and P.~J. Steinhardt, \emph{{Quintessence, cosmic
  coincidence, and the cosmological constant}},
  \href{https://doi.org/10.1103/PhysRevLett.82.896}{\emph{Phys. Rev. Lett.}
  {\bfseries 82} (1999) 896}
  [\href{https://arxiv.org/abs/astro-ph/9807002}{{\ttfamily
  astro-ph/9807002}}].

\bibitem{Steinhardt:1999nw}
P.~J. Steinhardt, L.-M. Wang and I.~Zlatev, \emph{{Cosmological tracking
  solutions}}, \href{https://doi.org/10.1103/PhysRevD.59.123504}{\emph{Phys.
  Rev. D} {\bfseries 59} (1999) 123504}
  [\href{https://arxiv.org/abs/astro-ph/9812313}{{\ttfamily
  astro-ph/9812313}}].

\bibitem{Sahni:1999gb}
V.~Sahni and A.~A. Starobinsky, \emph{{The Case for a positive cosmological
  Lambda term}}, \href{https://doi.org/10.1142/S0218271800000542}{\emph{Int. J.
  Mod. Phys. D} {\bfseries 9} (2000) 373}
  [\href{https://arxiv.org/abs/astro-ph/9904398}{{\ttfamily
  astro-ph/9904398}}].

\bibitem{Copeland:2006wr}
E.~J. Copeland, M.~Sami and S.~Tsujikawa, \emph{{Dynamics of dark energy}},
  \href{https://doi.org/10.1142/S021827180600942X}{\emph{Int. J. Mod. Phys. D.}
  {\bfseries 15} (2006) 1753}
  [\href{https://arxiv.org/abs/hep-th/0603057}{{\ttfamily hep-th/0603057}}].

\bibitem{Padmanabhan:2007xy}
T.~Padmanabhan, \emph{{Dark energy and gravity}},
  \href{https://doi.org/10.1007/s10714-007-0555-7}{\emph{Gen. Rel. Grav.}
  {\bfseries 40} (2008) 529} [\href{https://arxiv.org/abs/0705.2533}{{\ttfamily
  0705.2533}}].

\bibitem{Li:2011sd}
M.~Li, X.-D. Li, S.~Wang and Y.~Wang, \emph{{Dark Energy}},
  \href{https://doi.org/10.1088/0253-6102/56/3/24}{\emph{Commun. Theor. Phys.}
  {\bfseries 56} (2011) 525} [\href{https://arxiv.org/abs/1103.5870}{{\ttfamily
  1103.5870}}].

\bibitem{Benetti:2019gmo}
M.~Benetti and S.~Capozziello, \emph{{Connecting early and late epochs by
  f(z)CDM cosmography}},
  \href{https://doi.org/10.1088/1475-7516/2019/12/008}{\emph{JCAP} {\bfseries
  1912} (2019) 008} [\href{https://arxiv.org/abs/1910.09975}{{\ttfamily
  1910.09975}}].

\bibitem{Aghanim:2018eyx}
{\scshape Planck} collaboration, \emph{{Planck 2018 results. VI. Cosmological
  parameters}},  \href{https://arxiv.org/abs/1807.06209}{{\ttfamily
  1807.06209}}.

\bibitem{Riess:2019cxk}
A.~G. Riess, S.~Casertano, W.~Yuan, L.~M. Macri and D.~Scolnic, \emph{{Large
  Magellanic Cloud Cepheid Standards Provide a 1\% Foundation for the
  Determination of the Hubble Constant and Stronger Evidence for Physics beyond
  $\Lambda$CDM}},
  \href{https://doi.org/10.3847/1538-4357/ab1422}{\emph{Astrophys. J.}
  {\bfseries 876} (2019) 85}
  [\href{https://arxiv.org/abs/1903.07603}{{\ttfamily 1903.07603}}].

\bibitem{Schutz:1986gp}
B.~F. Schutz, \emph{{Determining the Hubble Constant from Gravitational Wave
  Observations}}, \href{https://doi.org/10.1038/323310a0}{\emph{Nature}
  {\bfseries 323} (1986) 310}.

\bibitem{Abbott:2017xzu}
{\scshape LIGO Scientific, Virgo, 1M2H, Dark Energy Camera GW-E, DES, DLT40,
  Las Cumbres Observatory, VINROUGE, MASTER} collaboration, \emph{{A
  gravitational-wave standard siren measurement of the Hubble constant}},
  \href{https://doi.org/10.1038/nature24471}{\emph{Nature} {\bfseries 551}
  (2017) 85} [\href{https://arxiv.org/abs/1710.05835}{{\ttfamily 1710.05835}}].

\bibitem{Yu:2017iju}
H.~Yu, B.~Ratra and F.-Y. Wang, \emph{{Hubble Parameter and Baryon Acoustic
  Oscillation Measurement Constraints on the Hubble Constant, the Deviation
  from the Spatially Flat $\Lambda$CDM Model, the Deceleration–Acceleration
  Transition Redshift, and Spatial Curvature}},
  \href{https://doi.org/10.3847/1538-4357/aab0a2}{\emph{Astrophys. J.}
  {\bfseries 856} (2018) 3} [\href{https://arxiv.org/abs/1711.03437}{{\ttfamily
  1711.03437}}].

\bibitem{Scolnic:2017caz}
D.~M. Scolnic et~al., \emph{{The Complete Light-curve Sample of
  Spectroscopically Confirmed SNe Ia from Pan-STARRS1 and Cosmological
  Constraints from the Combined Pantheon Sample}},
  \href{https://doi.org/10.3847/1538-4357/aab9bb}{\emph{Astrophys. J.}
  {\bfseries 859} (2018) 101}
  [\href{https://arxiv.org/abs/1710.00845}{{\ttfamily 1710.00845}}].

\bibitem{Riess:2017lxs}
A.~G. Riess et~al., \emph{{Type Ia Supernova Distances at Redshift $>$ 1.5 from
  the Hubble Space Telescope Multi-cycle Treasury Programs: The Early Expansion
  Rate}}, \href{https://doi.org/10.3847/1538-4357/aaa5a9}{\emph{Astrophys. J.}
  {\bfseries 853} (2018) 126}
  [\href{https://arxiv.org/abs/1710.00844}{{\ttfamily 1710.00844}}].

\bibitem{Gomez-Valent:2018hwc}
A.~Gómez-Valent and L.~Amendola, \emph{{$H_0$ from cosmic chronometers and
  Type Ia supernovae, with Gaussian Processes and the novel Weighted Polynomial
  Regression method}},
  \href{https://doi.org/10.1088/1475-7516/2018/04/051}{\emph{JCAP} {\bfseries
  1804} (2018) 051} [\href{https://arxiv.org/abs/1802.01505}{{\ttfamily
  1802.01505}}].

\bibitem{Haridasu:2018gqm}
B.~S. Haridasu, V.~V. Luković, M.~Moresco and N.~Vittorio, \emph{{An improved
  model-independent assessment of the late-time cosmic expansion}},
  \href{https://doi.org/10.1088/1475-7516/2018/10/015}{\emph{JCAP} {\bfseries
  1810} (2018) 015} [\href{https://arxiv.org/abs/1805.03595}{{\ttfamily
  1805.03595}}].

\bibitem{Gao:2013pfa}
Q.~Gao and Y.~Gong, \emph{{The tension on the cosmological parameters from
  different observational data}},
  \href{https://doi.org/10.1088/0264-9381/31/10/105007}{\emph{Class. Quant.
  Grav.} {\bfseries 31} (2014) 105007}
  [\href{https://arxiv.org/abs/1308.5627}{{\ttfamily 1308.5627}}].

\bibitem{Clarkson:2007pz}
C.~Clarkson, B.~Bassett and T.~H.-C. Lu, \emph{{A general test of the
  Copernican Principle}},
  \href{https://doi.org/10.1103/PhysRevLett.101.011301}{\emph{Phys. Rev. Lett.}
  {\bfseries 101} (2008) 011301}
  [\href{https://arxiv.org/abs/0712.3457}{{\ttfamily 0712.3457}}].

\bibitem{Sahni:2008xx}
V.~Sahni, A.~Shafieloo and A.~A. Starobinsky, \emph{{Two new diagnostics of
  dark energy}}, \href{https://doi.org/10.1103/PhysRevD.78.103502}{\emph{Phys.
  Rev. D.} {\bfseries 78} (2008) 103502}
  [\href{https://arxiv.org/abs/0807.3548}{{\ttfamily 0807.3548}}].

\bibitem{Zunckel:2008ti}
C.~Zunckel and C.~Clarkson, \emph{{Consistency Tests for the Cosmological
  Constant}}, \href{https://doi.org/10.1103/PhysRevLett.101.181301}{\emph{Phys.
  Rev. Lett.} {\bfseries 101} (2008) 181301}
  [\href{https://arxiv.org/abs/0807.4304}{{\ttfamily 0807.4304}}].

\bibitem{Nesseris:2010ep}
S.~Nesseris and A.~Shafieloo, \emph{{A model independent null test on the
  cosmological constant}},
  \href{https://doi.org/10.1111/j.1365-2966.2010.17254.x}{\emph{Mon. Not. Roy.
  Astron. Soc.} {\bfseries 408} (2010) 1879}
  [\href{https://arxiv.org/abs/1004.0960}{{\ttfamily 1004.0960}}].

\bibitem{Shafieloo:2012rs}
A.~Shafieloo, V.~Sahni and A.~A. Starobinsky, \emph{{A new null diagnostic
  customized for reconstructing the properties of dark energy from BAO data}},
  \href{https://doi.org/10.1103/PhysRevD.86.103527}{\emph{Phys. Rev. D}
  {\bfseries 86} (2012) 103527}
  [\href{https://arxiv.org/abs/1205.2870}{{\ttfamily 1205.2870}}].

\bibitem{Yahya:2013xma}
S.~Yahya, M.~Seikel, C.~Clarkson, R.~Maartens and M.~Smith, \emph{{Null tests
  of the cosmological constant using supernovae}},
  \href{https://doi.org/10.1103/PhysRevD.89.023503}{\emph{Phys. Rev.}
  {\bfseries D89} (2014) 023503}
  [\href{https://arxiv.org/abs/1308.4099}{{\ttfamily 1308.4099}}].

\bibitem{Nesseris:2014mfa}
S.~Nesseris and D.~Sapone, \emph{{Novel null-test for the $\Lambda$ cold dark
  matter model with growth-rate data}},
  \href{https://doi.org/10.1142/S0218271815500455}{\emph{Int. J. Mod. Phys. D}
  {\bfseries 24} (2015) 1550045}
  [\href{https://arxiv.org/abs/1409.3697}{{\ttfamily 1409.3697}}].

\bibitem{Marra:2017pst}
V.~Marra and D.~Sapone, \emph{{Null tests of the standard model using the
  linear model formalism}},
  \href{https://doi.org/10.1103/PhysRevD.97.083510}{\emph{Phys. Rev. D}
  {\bfseries 97} (2018) 083510}
  [\href{https://arxiv.org/abs/1712.09676}{{\ttfamily 1712.09676}}].

\bibitem{Franco:2019wbj}
F.~O. Franco, C.~Bonvin and C.~Clarkson, \emph{{A null test to probe the
  scale-dependence of the growth of structure as a test of General
  Relativity}}, \href{https://doi.org/10.1093/mnrasl/slz175}{\emph{Mon. Not.
  Roy. Astron. Soc.} {\bfseries 492} (2020) L34}
  [\href{https://arxiv.org/abs/1906.02217}{{\ttfamily 1906.02217}}].

\bibitem{Sahni:2014ooa}
V.~Sahni, A.~Shafieloo and A.~A. Starobinsky, \emph{{Model independent evidence
  for dark energy evolution from Baryon Acoustic Oscillations}},
  \href{https://doi.org/10.1088/2041-8205/793/2/L40}{\emph{Astrophys. J.}
  {\bfseries 793} (2014) L40}
  [\href{https://arxiv.org/abs/1406.2209}{{\ttfamily 1406.2209}}].

\bibitem{Clarkson:2010bm}
C.~Clarkson and C.~Zunckel, \emph{{Direct reconstruction of dark energy}},
  \href{https://doi.org/10.1103/PhysRevLett.104.211301}{\emph{Phys. Rev. Lett.}
  {\bfseries 104} (2010) 211301}
  [\href{https://arxiv.org/abs/1002.5004}{{\ttfamily 1002.5004}}].

\bibitem{Shafieloo:2012ht}
A.~Shafieloo, A.~G. Kim and E.~V. Linder, \emph{{Gaussian Process
  Cosmography}}, \href{https://doi.org/10.1103/PhysRevD.85.123530}{\emph{Phys.
  Rev. D} {\bfseries 85} (2012) 123530}
  [\href{https://arxiv.org/abs/1204.2272}{{\ttfamily 1204.2272}}].

\bibitem{Holsclaw:2010nb}
T.~Holsclaw, U.~Alam, B.~Sanso, H.~Lee, K.~Heitmann, S.~Habib et~al.,
  \emph{{Nonparametric Reconstruction of the Dark Energy Equation of State}},
  \href{https://doi.org/10.1103/PhysRevD.82.103502}{\emph{Phys. Rev. D}
  {\bfseries 82} (2010) 103502}
  [\href{https://arxiv.org/abs/1009.5443}{{\ttfamily 1009.5443}}].

\bibitem{Holsclaw:2010sk}
T.~Holsclaw, U.~Alam, B.~Sanso, H.~Lee, K.~Heitmann, S.~Habib et~al.,
  \emph{{Nonparametric Dark Energy Reconstruction from Supernova Data}},
  \href{https://doi.org/10.1103/PhysRevLett.105.241302}{\emph{Phys. Rev. Lett.}
  {\bfseries 105} (2010) 241302}
  [\href{https://arxiv.org/abs/1011.3079}{{\ttfamily 1011.3079}}].

\bibitem{Holsclaw:2011wi}
T.~Holsclaw, U.~Alam, B.~Sanso, H.~Lee, K.~Heitmann, S.~Habib et~al.,
  \emph{{Nonparametric Reconstruction of the Dark Energy Equation of State from
  Diverse Data Sets}},
  \href{https://doi.org/10.1103/PhysRevD.84.083501}{\emph{Phys. Rev. D}
  {\bfseries 84} (2011) 083501}
  [\href{https://arxiv.org/abs/1104.2041}{{\ttfamily 1104.2041}}].

\bibitem{Bilicki:2012ub}
M.~Bilicki and M.~Seikel, \emph{{We do not live in the $R_h = c t$ universe}},
  \href{https://doi.org/10.1111/j.1365-2966.2012.21575.x}{\emph{Mon. Not. Roy.
  Astron. Soc.} {\bfseries 425} (2012) 1664}
  [\href{https://arxiv.org/abs/1206.5130}{{\ttfamily 1206.5130}}].

\bibitem{Seikel:2012uu}
M.~Seikel, C.~Clarkson and M.~Smith, \emph{{Reconstruction of dark energy and
  expansion dynamics using Gaussian processes}},
  \href{https://doi.org/10.1088/1475-7516/2012/06/036}{\emph{JCAP} {\bfseries
  1206} (2012) 036} [\href{https://arxiv.org/abs/1204.2832}{{\ttfamily
  1204.2832}}].

\bibitem{Seikel:2012cs}
M.~Seikel, S.~Yahya, R.~Maartens and C.~Clarkson, \emph{{Using H(z) data as a
  probe of the concordance model}},
  \href{https://doi.org/10.1103/PhysRevD.86.083001}{\emph{Phys. Rev.}
  {\bfseries D86} (2012) 083001}
  [\href{https://arxiv.org/abs/1205.3431}{{\ttfamily 1205.3431}}].

\bibitem{Seikel:2013fda}
M.~Seikel and C.~Clarkson, \emph{{Optimising Gaussian processes for
  reconstructing dark energy dynamics from supernovae}},
  \href{https://arxiv.org/abs/1311.6678}{{\ttfamily 1311.6678}}.

\bibitem{Nair:2013sna}
R.~Nair, S.~Jhingan and D.~Jain, \emph{{Exploring scalar field dynamics with
  Gaussian processes}},
  \href{https://doi.org/10.1088/1475-7516/2014/01/005}{\emph{JCAP} {\bfseries
  1401} (2014) 005} [\href{https://arxiv.org/abs/1306.0606}{{\ttfamily
  1306.0606}}].

\bibitem{Busti:2014dua}
V.~C. Busti, C.~Clarkson and M.~Seikel, \emph{{Evidence for a Lower Value for
  $H_0$ from Cosmic Chronometers Data?}},
  \href{https://doi.org/10.1093/mnrasl/slu035}{\emph{Mon. Not. Roy. Astron.
  Soc.} {\bfseries 441} (2014) 11}
  [\href{https://arxiv.org/abs/1402.5429}{{\ttfamily 1402.5429}}].

\bibitem{Verde:2014qea}
L.~Verde, P.~Protopapas and R.~Jimenez, \emph{{The expansion rate of the
  intermediate Universe in light of Planck}},
  \href{https://doi.org/10.1016/j.dark.2014.09.003}{\emph{Phys. Dark Univ.}
  {\bfseries 5-6} (2014) 307}
  [\href{https://arxiv.org/abs/1403.2181}{{\ttfamily 1403.2181}}].

\bibitem{Li:2015nta}
Z.~Li, J.~E. Gonzalez, H.~Yu, Z.-H. Zhu and J.~S. Alcaniz, \emph{{Constructing
  a cosmological model-independent Hubble diagram of type Ia supernovae with
  cosmic chronometers}},
  \href{https://doi.org/10.1103/PhysRevD.93.043014}{\emph{Phys. Rev. D}
  {\bfseries 93} (2016) 043014}
  [\href{https://arxiv.org/abs/1504.03269}{{\ttfamily 1504.03269}}].

\bibitem{Vitenti:2015aaa}
S.~D.~P. Vitenti and M.~Penna-Lima, \emph{{A general reconstruction of the
  recent expansion history of the universe}},
  \href{https://doi.org/10.1088/1475-7516/2015/9/045,
  10.1088/1475-7516/2015/09/045}{\emph{JCAP} {\bfseries 1509} (2015) 045}
  [\href{https://arxiv.org/abs/1505.01883}{{\ttfamily 1505.01883}}].

\bibitem{Wang:2016iij}
D.~Wang and X.-H. Meng, \emph{{Model-independent determination on H$_{0}$ using
  the latest cosmic chronometer data}},
  \href{https://doi.org/10.1007/s11433-017-9079-1}{\emph{Sci. China Phys. Mech.
  Astron.} {\bfseries 60} (2017) 110411}
  [\href{https://arxiv.org/abs/1610.01202}{{\ttfamily 1610.01202}}].

\bibitem{Zhang:2016tto}
M.-J. Zhang and J.-Q. Xia, \emph{{Test of the cosmic evolution using Gaussian
  processes}}, \href{https://doi.org/10.1088/1475-7516/2016/12/005}{\emph{JCAP}
  {\bfseries 1612} (2016) 005}
  [\href{https://arxiv.org/abs/1606.04398}{{\ttfamily 1606.04398}}].

\bibitem{Wei:2016xti}
J.-J. Wei and X.-F. Wu, \emph{{An Improved Method to Measure the Cosmic
  Curvature}}, \href{https://doi.org/10.3847/1538-4357/aa674b}{\emph{Astrophys.
  J.} {\bfseries 838} (2017) 160}
  [\href{https://arxiv.org/abs/1611.00904}{{\ttfamily 1611.00904}}].

\bibitem{Yennapureddy:2017vvb}
M.~K. Yennapureddy and F.~Melia, \emph{{Reconstruction of the HII Galaxy Hubble
  Diagram using Gaussian Processes}},
  \href{https://doi.org/10.1088/1475-7516/2017/11/029}{\emph{JCAP} {\bfseries
  1711} (2017) 029} [\href{https://arxiv.org/abs/1711.03454}{{\ttfamily
  1711.03454}}].

\bibitem{Melia:2018tzi}
F.~Melia and M.~K. Yennapureddy, \emph{{Model Selection Using Cosmic
  Chronometers with Gaussian Processes}},
  \href{https://doi.org/10.1088/1475-7516/2018/02/034}{\emph{JCAP} {\bfseries
  1802} (2018) 034} [\href{https://arxiv.org/abs/1802.02255}{{\ttfamily
  1802.02255}}].

\bibitem{Pinho:2018unz}
A.~M. Pinho, S.~Casas and L.~Amendola, \emph{{Model-independent reconstruction
  of the linear anisotropic stress $\eta$}},
  \href{https://doi.org/10.1088/1475-7516/2018/11/027}{\emph{JCAP} {\bfseries
  1811} (2018) 027} [\href{https://arxiv.org/abs/1805.00027}{{\ttfamily
  1805.00027}}].

\bibitem{Jesus:2019nnk}
J.~F. Jesus, R.~Valentim, A.~A. Escobal and S.~H. Pereira, \emph{{Gaussian
  Process Estimation of Transition Redshift}},
  \href{https://arxiv.org/abs/1909.00090}{{\ttfamily 1909.00090}}.

\bibitem{Bengaly:2019ibu}
C.~A. Bengaly, \emph{{Evidence for cosmic acceleration with next-generation
  surveys: A model-independent approach}},
  \href{https://doi.org/10.1093/mnrasl/slaa040}{\emph{Mon. Not. Roy. Astron.
  Soc.} (2020) in press} [\href{https://arxiv.org/abs/1912.05528}{{\ttfamily
  1912.05528}}].

\bibitem{John:1999gm}
M.~V. John and K.~B. Joseph, \emph{{Generalized Chen-Wu type cosmological
  model}}, \href{https://doi.org/10.1103/PhysRevD.61.087304}{\emph{Phys. Rev.
  D} {\bfseries 61} (2000) 087304}
  [\href{https://arxiv.org/abs/gr-qc/9912069}{{\ttfamily gr-qc/9912069}}].

\bibitem{Melia:2007sd}
F.~Melia, \emph{{The Cosmic Horizon}},
  \href{https://doi.org/10.1111/j.1365-2966.2007.12499.x}{\emph{Mon. Not. Roy.
  Astron. Soc.} {\bfseries 382} (2007) 1917}
  [\href{https://arxiv.org/abs/0711.4181}{{\ttfamily 0711.4181}}].

\bibitem{Melia:2011fj}
F.~Melia and A.~Shevchuk, \emph{{The $R_h = ct$ Universe}},
  \href{https://doi.org/10.1111/j.1365-2966.2011.19906.x}{\emph{Mon. Not. Roy.
  Astron. Soc.} {\bfseries 419} (2012) 2579}
  [\href{https://arxiv.org/abs/1109.5189}{{\ttfamily 1109.5189}}].

\bibitem{Lopez-Corredoira:2016pwg}
M.~Lopez-Corredoira, F.~Melia, E.~Lusso and G.~Risaliti, \emph{{Cosmological
  test with the QSO Hubble diagram}},
  \href{https://doi.org/10.1142/S0218271816500607}{\emph{Int. J. Mod. Phys. D}
  {\bfseries 25} (2016) 1650060}
  [\href{https://arxiv.org/abs/1602.06743}{{\ttfamily 1602.06743}}].

\bibitem{Melia:2016djn}
F.~Melia, \emph{{The Linear Growth of Structure in the $R_h=ct$ Universe}},
  \href{https://doi.org/10.1093/mnras/stw2493}{\emph{Mon. Not. Roy. Astron.
  Soc.} {\bfseries 464} (2017) 1966}
  [\href{https://arxiv.org/abs/1609.08576}{{\ttfamily 1609.08576}}].

\bibitem{Melia:2018nfw}
F.~Melia, \emph{{A comparison of the Rh = ct and $\Lambda$CDM cosmologies using
  the cosmic distance duality relation}},
  \href{https://doi.org/10.1093/mnras/sty2596}{\emph{Mon. Not. Roy. Astron.
  Soc.} {\bfseries 481} (2018) 4855}
  [\href{https://arxiv.org/abs/1804.09906}{{\ttfamily 1804.09906}}].

\bibitem{John:2019nlw}
M.~V. John, \emph{{$R_h=ct$ and the eternal coasting cosmological model}},
  \href{https://doi.org/10.1093/mnrasl/sly243}{\emph{Mon. Not. Roy. Astron.
  Soc.} {\bfseries 484} (2019) L35}
  [\href{https://arxiv.org/abs/1902.05088}{{\ttfamily 1902.05088}}].

\bibitem{Yennapureddy:2019omi}
M.~K. Yennapureddy and F.~Melia, \emph{{A comparison of the $R_{\mathrm{h}}=ct$
  and $\varLambda$CDM cosmologies based on the observed halo mass function}},
  \href{https://doi.org/10.1140/epjc/s10052-019-7082-z}{\emph{Eur. Phys. J. C}
  {\bfseries 79} (2019) 571}
  [\href{https://arxiv.org/abs/1907.00897}{{\ttfamily 1907.00897}}].

\bibitem{Capozziello:2018jya}
S.~Capozziello, Ruchika and A.~A. Sen, \emph{{Model independent constraints on
  dark energy evolution from low-redshift observations}},
  \href{https://doi.org/10.1093/mnras/stz176}{\emph{Mon. Not. Roy. Astron.
  Soc.} {\bfseries 484} (2019) 4484}
  [\href{https://arxiv.org/abs/1806.03943}{{\ttfamily 1806.03943}}].

\bibitem{Arjona:2019fwb}
R.~Arjona and S.~Nesseris, \emph{{What can Machine Learning tell us about the
  background expansion of the Universe?}},
  \href{https://arxiv.org/abs/1910.01529}{{\ttfamily 1910.01529}}.

\bibitem{Visser:1997qk}
M.~Visser, \emph{{Energy conditions in the epoch of galaxy formation}},
  \href{https://doi.org/10.1126/science.276.5309.88}{\emph{Science} {\bfseries
  276} (1997) 88} [\href{https://arxiv.org/abs/1501.01619}{{\ttfamily
  1501.01619}}].

\bibitem{Visser:1997tq}
M.~Visser, \emph{{General relativistic energy conditions: The Hubble expansion
  in the epoch of galaxy formation}},
  \href{https://doi.org/10.1103/PhysRevD.56.7578}{\emph{Phys. Rev. D}
  {\bfseries 56} (1997) 7578}
  [\href{https://arxiv.org/abs/gr-qc/9705070}{{\ttfamily gr-qc/9705070}}].

\bibitem{Santos:2006ja}
J.~Santos, J.~S. Alcaniz and M.~J. Reboucas, \emph{{Energy Conditions and
  Supernovae Observations}},
  \href{https://doi.org/10.1103/PhysRevD.74.067301}{\emph{Phys. Rev. D}
  {\bfseries 74} (2006) 067301}
  [\href{https://arxiv.org/abs/astro-ph/0608031}{{\ttfamily
  astro-ph/0608031}}].

\bibitem{Santos:2007pp}
J.~Santos, J.~S. Alcaniz, N.~Pires and M.~J. Reboucas, \emph{{Energy Conditions
  and Cosmic Acceleration}},
  \href{https://doi.org/10.1103/PhysRevD.75.083523}{\emph{Phys. Rev. D}
  {\bfseries 75} (2007) 083523}
  [\href{https://arxiv.org/abs/astro-ph/0702728}{{\ttfamily
  astro-ph/0702728}}].

\bibitem{Gong:2007fm}
Y.~Gong and A.~Wang, \emph{{Energy conditions and current acceleration of the
  universe}}, \href{https://doi.org/10.1016/j.physletb.2007.06.065}{\emph{Phys.
  Lett. B} {\bfseries 652} (2007) 63}
  [\href{https://arxiv.org/abs/0705.0996}{{\ttfamily 0705.0996}}].

\bibitem{Gong:2007zf}
Y.~Gong, A.~Wang, Q.~Wu and Y.-Z. Zhang, \emph{{Direct evidence of acceleration
  from distance modulus redshift graph}},
  \href{https://doi.org/10.1088/1475-7516/2007/08/018}{\emph{JCAP} {\bfseries
  0708} (2007) 018} [\href{https://arxiv.org/abs/astro-ph/0703583}{{\ttfamily
  astro-ph/0703583}}].

\bibitem{Seikel:2007pk}
M.~Seikel and D.~J. Schwarz, \emph{{How strong is the evidence for accelerated
  expansion?}},
  \href{https://doi.org/10.1088/1475-7516/2008/02/007}{\emph{JCAP} {\bfseries
  0802} (2008) 007} [\href{https://arxiv.org/abs/0711.3180}{{\ttfamily
  0711.3180}}].

\bibitem{Velten:2017ire}
H.~Velten, S.~Gomes and V.~C. Busti, \emph{{Gauging the cosmic acceleration
  with recent type Ia supernovae data sets}},
  \href{https://doi.org/10.1103/PhysRevD.97.083516}{\emph{Phys. Rev.}
  {\bfseries D97} (2018) 083516}
  [\href{https://arxiv.org/abs/1801.00114}{{\ttfamily 1801.00114}}].

\bibitem{gong:2006tx}
Y.~Gong and A.~Wang, \emph{{Observational constraints on the acceleration of
  the universe}}, \href{https://doi.org/10.1103/PhysRevD.73.083506}{\emph{Phys.
  Rev. D} {\bfseries 73} (2006) 083506}
  [\href{https://arxiv.org/abs/astro-ph/0601453}{{\ttfamily
  astro-ph/0601453}}].

\bibitem{gong:2006gs}
Y.-G. Gong and A.~Wang, \emph{{Reconstruction of the deceleration parameter and
  the equation of state of dark energy}},
  \href{https://doi.org/10.1103/PhysRevD.75.043520}{\emph{Phys. Rev.}
  {\bfseries D75} (2007) 043520}
  [\href{https://arxiv.org/abs/astro-ph/0612196}{{\ttfamily
  astro-ph/0612196}}].

\bibitem{Gao:2012ef}
Q.~Gao and Y.~Gong, \emph{{Constraints on slow-roll thawing models from
  fundamental constants}},
  \href{https://doi.org/10.1142/S0218271813500351}{\emph{Int. J. Mod. Phys.}
  {\bfseries D22} (2013) 1350035}
  [\href{https://arxiv.org/abs/1212.6815}{{\ttfamily 1212.6815}}].

\bibitem{Gong:2013bn}
Y.~Gong and Q.~Gao, \emph{{On the effect of the degeneracy among dark energy
  parameters}},
  \href{https://doi.org/10.1140/epjc/s10052-014-2729-2}{\emph{Eur. Phys. J.}
  {\bfseries C74} (2014) 2729}
  [\href{https://arxiv.org/abs/1301.1224}{{\ttfamily 1301.1224}}].

\bibitem{Chevallier:2000qy}
M.~Chevallier and D.~Polarski, \emph{{Accelerating universes with scaling dark
  matter}}, \href{https://doi.org/10.1142/S0218271801000822}{\emph{Int. J. Mod.
  Phys. D} {\bfseries 10} (2001) 213}
  [\href{https://arxiv.org/abs/gr-qc/0009008}{{\ttfamily gr-qc/0009008}}].

\bibitem{Linder:2002et}
E.~V. Linder, \emph{{Exploring the expansion history of the universe}},
  \href{https://doi.org/10.1103/PhysRevLett.90.091301}{\emph{Phys. Rev. Lett.}
  {\bfseries 90} (2003) 091301}
  [\href{https://arxiv.org/abs/astro-ph/0208512}{{\ttfamily
  astro-ph/0208512}}].

\bibitem{Jimenez:2001gg}
R.~Jimenez and A.~Loeb, \emph{{Constraining cosmological parameters based on
  relative galaxy ages}},
  \href{https://doi.org/10.1086/340549}{\emph{Astrophys. J.} {\bfseries 573}
  (2002) 37} [\href{https://arxiv.org/abs/astro-ph/0106145}{{\ttfamily
  astro-ph/0106145}}].

\bibitem{Farooq:2016zwm}
O.~Farooq, F.~R. Madiyar, S.~Crandall and B.~Ratra, \emph{{Hubble Parameter
  Measurement Constraints on the Redshift of the Deceleration–acceleration
  Transition, Dynamical Dark Energy, and Space Curvature}},
  \href{https://doi.org/10.3847/1538-4357/835/1/26}{\emph{Astrophys. J.}
  {\bfseries 835} (2017) 26}
  [\href{https://arxiv.org/abs/1607.03537}{{\ttfamily 1607.03537}}].

\bibitem{Gomez-Valent:2018gvm}
A.~Gómez-Valent, \emph{{Quantifying the evidence for the current speed-up of
  the Universe with low and intermediate-redshift data. A more
  model-independent approach}},
  \href{https://doi.org/10.1088/1475-7516/2019/05/026}{\emph{JCAP} {\bfseries
  1905} (2019) 026} [\href{https://arxiv.org/abs/1810.02278}{{\ttfamily
  1810.02278}}].

\bibitem{Bruzual:2003tq}
G.~Bruzual and S.~Charlot, \emph{{Stellar population synthesis at the
  resolution of 2003}},
  \href{https://doi.org/10.1046/j.1365-8711.2003.06897.x}{\emph{Mon. Not. Roy.
  Astron. Soc.} {\bfseries 344} (2003) 1000}
  [\href{https://arxiv.org/abs/astro-ph/0309134}{{\ttfamily
  astro-ph/0309134}}].

\bibitem{Maraston:2011sq}
C.~Maraston and G.~Stromback, \emph{{Stellar population models at high spectral
  resolution}},
  \href{https://doi.org/10.1111/j.1365-2966.2011.19738.x}{\emph{Mon. Not. Roy.
  Astron. Soc.} {\bfseries 418} (2011) 2785}
  [\href{https://arxiv.org/abs/1109.0543}{{\ttfamily 1109.0543}}].

\bibitem{Zhang:2012mp}
C.~Zhang, H.~Zhang, S.~Yuan, T.-J. Zhang and Y.-C. Sun, \emph{{Four new
  observational $H(z)$ data from luminous red galaxies in the Sloan Digital Sky
  Survey data release seven}},
  \href{https://doi.org/10.1088/1674-4527/14/10/002}{\emph{Res. Astron.
  Astrophys.} {\bfseries 14} (2014) 1221}
  [\href{https://arxiv.org/abs/1207.4541}{{\ttfamily 1207.4541}}].

\bibitem{Moresco:2016mzx}
M.~Moresco, L.~Pozzetti, A.~Cimatti, R.~Jimenez, C.~Maraston, L.~Verde et~al.,
  \emph{{A 6\% measurement of the Hubble parameter at $z\sim0.45$: direct
  evidence of the epoch of cosmic re-acceleration}},
  \href{https://doi.org/10.1088/1475-7516/2016/05/014}{\emph{JCAP} {\bfseries
  1605} (2016) 014} [\href{https://arxiv.org/abs/1601.01701}{{\ttfamily
  1601.01701}}].

\bibitem{Simon:2004tf}
J.~Simon, L.~Verde and R.~Jimenez, \emph{{Constraints on the redshift
  dependence of the dark energy potential}},
  \href{https://doi.org/10.1103/PhysRevD.71.123001}{\emph{Phys. Rev.}
  {\bfseries D71} (2005) 123001}
  [\href{https://arxiv.org/abs/astro-ph/0412269}{{\ttfamily
  astro-ph/0412269}}].

\bibitem{Stern:2009ep}
D.~Stern, R.~Jimenez, L.~Verde, M.~Kamionkowski and S.~A. Stanford,
  \emph{{Cosmic Chronometers: Constraining the Equation of State of Dark
  Energy. I: H(z) Measurements}},
  \href{https://doi.org/10.1088/1475-7516/2010/02/008}{\emph{JCAP} {\bfseries
  1002} (2010) 008} [\href{https://arxiv.org/abs/0907.3149}{{\ttfamily
  0907.3149}}].

\bibitem{Moresco:2012jh}
M.~Moresco et~al., \emph{{Improved constraints on the expansion rate of the
  Universe up to z~1.1 from the spectroscopic evolution of cosmic
  chronometers}},
  \href{https://doi.org/10.1088/1475-7516/2012/08/006}{\emph{JCAP} {\bfseries
  1208} (2012) 006} [\href{https://arxiv.org/abs/1201.3609}{{\ttfamily
  1201.3609}}].

\bibitem{Moresco:2015cya}
M.~Moresco, \emph{{Raising the bar: new constraints on the Hubble parameter
  with cosmic chronometers at z$\sim$2}},
  \href{https://doi.org/10.1093/mnrasl/slv037}{\emph{Mon. Not. Roy. Astron.
  Soc.} {\bfseries 450} (2015) L16}
  [\href{https://arxiv.org/abs/1503.01116}{{\ttfamily 1503.01116}}].

\bibitem{Ratsimbazafy:2017vga}
A.~L. Ratsimbazafy, S.~I. Loubser, S.~M. Crawford, C.~M. Cress, B.~A. Bassett,
  R.~C. Nichol et~al., \emph{{Age-dating Luminous Red Galaxies observed with
  the Southern African Large Telescope}},
  \href{https://doi.org/10.1093/mnras/stx301}{\emph{Mon. Not. Roy. Astron.
  Soc.} {\bfseries 467} (2017) 3239}
  [\href{https://arxiv.org/abs/1702.00418}{{\ttfamily 1702.00418}}].

\bibitem{Rest:2013mwz}
A.~Rest et~al., \emph{{Cosmological Constraints from Measurements of Type Ia
  Supernovae discovered during the first 1.5 yr of the Pan-STARRS1 Survey}},
  \href{https://doi.org/10.1088/0004-637X/795/1/44}{\emph{Astrophys. J.}
  {\bfseries 795} (2014) 44} [\href{https://arxiv.org/abs/1310.3828}{{\ttfamily
  1310.3828}}].

\bibitem{Hicken:2009df}
M.~Hicken, P.~Challis, S.~Jha, R.~P. Kirsher, T.~Matheson, M.~Modjaz et~al.,
  \emph{{CfA3: 185 Type Ia Supernova Light Curves from the CfA}},
  \href{https://doi.org/10.1088/0004-637X/700/1/331}{\emph{Astrophys. J.}
  {\bfseries 700} (2009) 331}
  [\href{https://arxiv.org/abs/0901.4787}{{\ttfamily 0901.4787}}].

\bibitem{Stritzinger:2011qd}
M.~D. Stritzinger et~al., \emph{{The Carnegie Supernova Project: Second
  Photometry Data Release of Low-Redshift Type Ia Supernovae}},
  \href{https://doi.org/10.1088/0004-6256/142/5/156}{\emph{Astron. J.}
  {\bfseries 142} (2011) 156}
  [\href{https://arxiv.org/abs/1108.3108}{{\ttfamily 1108.3108}}].

\bibitem{Kessler:2009ys}
R.~Kessler et~al., \emph{{First-year Sloan Digital Sky Survey-II (SDSS-II)
  Supernova Results: Hubble Diagram and Cosmological Parameters}},
  \href{https://doi.org/10.1088/0067-0049/185/1/32}{\emph{Astrophys. J. Suppl.}
  {\bfseries 185} (2009) 32} [\href{https://arxiv.org/abs/0908.4274}{{\ttfamily
  0908.4274}}].

\bibitem{Conley:2011ku}
{\scshape SNLS} collaboration, \emph{{Supernova Constraints and Systematic
  Uncertainties from the First 3 Years of the Supernova Legacy Survey}},
  \href{https://doi.org/10.1088/0067-0049/192/1/1}{\emph{Astrophys. J. Suppl.}
  {\bfseries 192} (2011) 1} [\href{https://arxiv.org/abs/1104.1443}{{\ttfamily
  1104.1443}}].

\bibitem{Suzuki:2011hu}
{\scshape Supernova Cosmology Project} collaboration, \emph{{The Hubble Space
  Telescope Cluster Supernova Survey: V. Improving the Dark Energy Constraints
  Above $z>1$ and Building an Early-Type-Hosted Supernova Sample}},
  \href{https://doi.org/10.1088/0004-637X/746/1/85}{\emph{Astrophys. J.}
  {\bfseries 746} (2012) 85} [\href{https://arxiv.org/abs/1105.3470}{{\ttfamily
  1105.3470}}].

\bibitem{Riess:2006fw}
A.~G. Riess et~al., \emph{{New Hubble Space Telescope Discoveries of Type Ia
  Supernovae at $z>=1$: Narrowing Constraints on the Early Behavior of Dark
  Energy}}, \href{https://doi.org/10.1086/510378}{\emph{Astrophys. J.}
  {\bfseries 659} (2007) 98}
  [\href{https://arxiv.org/abs/astro-ph/0611572}{{\ttfamily
  astro-ph/0611572}}].

\bibitem{Rodney:2014twa}
S.~A. Rodney et~al., \emph{{Type Ia Supernova Rate Measurements to Redshift 2.5
  from CANDELS : Searching for Prompt Explosions in the Early Universe}},
  \href{https://doi.org/10.1088/0004-6256/148/1/13}{\emph{Astron. J.}
  {\bfseries 148} (2014) 13} [\href{https://arxiv.org/abs/1401.7978}{{\ttfamily
  1401.7978}}].

\bibitem{Graur:2013msa}
O.~Graur et~al., \emph{{Type-Ia Supernova Rates to Redshift 2.4 from CLASH: the
  Cluster Lensing And Supernova survey with Hubble}},
  \href{https://doi.org/10.1088/0004-637X/783/1/28}{\emph{Astrophys. J.}
  {\bfseries 783} (2014) 28} [\href{https://arxiv.org/abs/1310.3495}{{\ttfamily
  1310.3495}}].

\bibitem{Evslin:2017qdn}
J.~Evslin, A.~A. Sen and Ruchika, \emph{{Price of shifting the Hubble
  constant}}, \href{https://doi.org/10.1103/PhysRevD.97.103511}{\emph{Phys.
  Rev. D} {\bfseries 97} (2018) 103511}
  [\href{https://arxiv.org/abs/1711.01051}{{\ttfamily 1711.01051}}].

\bibitem{Beutler:2011hx}
F.~Beutler, C.~Blake, M.~Colless, D.~H. Jones, L.~Staveley-Smith, L.~Campbell
  et~al., \emph{{The 6dF Galaxy Survey: Baryon Acoustic Oscillations and the
  Local Hubble Constant}},
  \href{https://doi.org/10.1111/j.1365-2966.2011.19250.x}{\emph{Mon. Not. Roy.
  Astron. Soc.} {\bfseries 416} (2011) 3017}
  [\href{https://arxiv.org/abs/1106.3366}{{\ttfamily 1106.3366}}].

\bibitem{Ross:2014qpa}
A.~J. Ross, L.~Samushia, C.~Howlett, W.~J. Percival, A.~Burden and M.~Manera,
  \emph{{The clustering of the SDSS DR7 main Galaxy sample – I. A 4 per cent
  distance measure at $z = 0.15$}},
  \href{https://doi.org/10.1093/mnras/stv154}{\emph{Mon. Not. Roy. Astron.
  Soc.} {\bfseries 449} (2015) 835}
  [\href{https://arxiv.org/abs/1409.3242}{{\ttfamily 1409.3242}}].

\bibitem{Ata:2017dya}
M.~Ata et~al., \emph{{The clustering of the SDSS-IV extended Baryon Oscillation
  Spectroscopic Survey DR14 quasar sample: first measurement of baryon acoustic
  oscillations between redshift 0.8 and 2.2}},
  \href{https://doi.org/10.1093/mnras/stx2630}{\emph{Mon. Not. Roy. Astron.
  Soc.} {\bfseries 473} (2018) 4773}
  [\href{https://arxiv.org/abs/1705.06373}{{\ttfamily 1705.06373}}].

\bibitem{Alam:2016hwk}
{\scshape BOSS} collaboration, \emph{{The clustering of galaxies in the
  completed SDSS-III Baryon Oscillation Spectroscopic Survey: cosmological
  analysis of the DR12 galaxy sample}},
  \href{https://doi.org/10.1093/mnras/stx721}{\emph{Mon. Not. Roy. Astron.
  Soc.} {\bfseries 470} (2017) 2617}
  [\href{https://arxiv.org/abs/1607.03155}{{\ttfamily 1607.03155}}].

\bibitem{Bourboux:2017cbm}
H.~du~Mas~des Bourboux et~al., \emph{{Baryon acoustic oscillations from the
  complete SDSS-III Ly$\alpha$-quasar cross-correlation function at $z=2.4$}},
  \href{https://doi.org/10.1051/0004-6361/201731731}{\emph{Astron. Astrophys.}
  {\bfseries 608} (2017) A130}
  [\href{https://arxiv.org/abs/1708.02225}{{\ttfamily 1708.02225}}].

\bibitem{Eisenstein:2005su}
{\scshape SDSS} collaboration, \emph{{Detection of the Baryon Acoustic Peak in
  the Large-Scale Correlation Function of SDSS Luminous Red Galaxies}},
  \href{https://doi.org/10.1086/466512}{\emph{Astrophys. J.} {\bfseries 633}
  (2005) 560} [\href{https://arxiv.org/abs/astro-ph/0501171}{{\ttfamily
  astro-ph/0501171}}].

\bibitem{Eisenstein:1997ik}
D.~J. Eisenstein and W.~Hu, \emph{{Baryonic features in the matter transfer
  function}}, \href{https://doi.org/10.1086/305424}{\emph{Astrophys. J.}
  {\bfseries 496} (1998) 605}
  [\href{https://arxiv.org/abs/astro-ph/9709112}{{\ttfamily
  astro-ph/9709112}}].

\end{thebibliography}

\providecommand{\href}[2]{#2}\begingroup\raggedright\endgroup

\end{document}